\documentclass[preprint,12pt]{elsarticle}

\usepackage{amssymb}

\usepackage{setspace}
\usepackage{amsthm}
\theoremstyle{definition}

\usepackage{graphicx}
\usepackage{dcolumn}
\usepackage{multirow}
\usepackage{subfigure}
\usepackage{times,mathptm}
\usepackage{float}
\usepackage{color}
\usepackage{amsmath,amsfonts}
\usepackage{mathptmx}
\usepackage{mathrsfs}
\usepackage{bbm}
\usepackage{bm}
\usepackage{xfrac}

\newcommand{\beq}{\begin{equation}}
\newcommand{\eeq}{\end{equation}} 
\newcommand{\bea}{\begin{eqnarray}}
\newcommand{\eea}{\end{eqnarray}} 

\newcommand{\ih}{\hat{\imath}}

\newcommand{\ua}{\uparrow}
\newcommand{\da}{\downarrow}

\renewcommand{\d}{\delta}

\renewcommand{\l}{\lambda}

\newcommand{\pbar}{\overline{\psi}}

\renewcommand{\b}{\beta}
\renewcommand{\a}{\alpha}

\renewcommand{\ni}{\noindent}
\newcommand{\tr}{\text{Tr}}

\newcommand{\bx}{\mathbf{x}}

\newcommand{\vx}{{\vec{x}}}

\newcommand{\n}{\nu}
\newcommand{\m}{\mu}

\newcommand{\g}{\gamma}

\newcommand{\e}{\epsilon}
\newcommand{\s}{\sigma}
\renewcommand{\k}{\kappa}

\renewcommand{\th}{\theta}

\newcommand{\oh}{\frac{1}{2}}

\newcommand{\dg}{\dagger}
\newcommand{\non}{\nonumber}
\renewcommand{\t}{\tau}
\newcommand{\rf}[1]{(\ref{#1})}
\newcommand{\ra}{\rightarrow}
\newcommand{\pa}{\partial}
\renewcommand{\vec}[1]{\bm #1}

\journal{Annals of Physics}

\begin{document}

\begin{frontmatter}



\title{Cuprates and center vortices:  \\ A QCD confinement mechanism in a high-T$_\text{c}$ context}


\author[1]{Jeff Greensite}
\ead{greensit@sfsu.edu}
\author[1]{Kazue Matsuyama} 
\ead{kazuem@sfsu.edu}
\address[1]{Physics and Astronomy Department, San Francisco State University, \\ 1600 Holloway Ave,
San Francisco, CA 94132 USA}

\begin{abstract}
 
    It is suggested that the center vortex confinement mechanism, familiar in  hadronic physics, may have some
relevance to high-T$_\text{c}$ phenomena.  We focus specifically on the transition from the superconducting phase to
the pseudogap phase.  There is evidence of a vortex liquid in the latter phase, in which the pairing responsible for superconductivity still exists, but superconductivity itself does not.  An analogy, drawn from particle physics, may be the Higgs to confinement phase transition in an SU(N) gauge theory, where the confined phase is a vortex liquid, and the Higgs phase is a phase of a broken global 
$Z_N$ symmetry.  We illustrate this idea with numerical simulations of a spatially asymmetric U(1) gauge-Higgs model, with lattice artifact monopoles suppressed.  We show the existence of a Higgs (superconductor) to confinement (vortex liquid) phase, explicitly identifying vortices in lattice configurations generated in the confined phase, and showing that they produce an area-law falloff in planar Wilson loops, which may be measurable experimentally.   The superconducting phase is a phase of broken global
Z$_\text{2}$  symmetry. 
\end{abstract}


\end{frontmatter}




\bibliographystyle{elsarticle-num}

\section{\label{intro}Introduction}

    In a series of articles that appeared over a decade ago, Ong et al.\ \cite{2006PhRvB..73b4510W,2007JMMM..310..460L,2010PhRvB..81e4510L} presented evidence that the pseudogap phase in the cuprates behaves in some ways as a vortex liquid;  more recently the idea has been discussed by Anderson  \cite{2018arXiv180411186A,2016arXiv161203919A}.  Since the evidence presented by Ong et al.\  also suggests that the pairing responsible for superconductivity persists in the pseudogap region, the question is why superconductivity is absent in this region.   The answer given in the cited references (see also
\cite{Emery}) is that superconductivity is a ``phase locked'' region, where this expression refers to the phase of the order parameter, while the pseudogap region is characterized by spatial disorder in the phase of the order parameter, which is due to the existence of a disordered vortex liquid.  Of course a gauge choice, e.g.\ London gauge, is implicit in this picture, since the phase of the order parameter is a gauge-variant quantity.   We will argue here, in the context of an effective U(1) gauge-Higgs theory with a no-monopole constraint, that the vortex liquid and superconductor phases can be distinguished by the unbroken or spontaneously broken realization of a global Z$_\text{2}$ symmetry, and in the process we will make contact with one of the proposed mechanisms of quark confinement, known as the center vortex mechanism, in non-abelian gauge theories.

    In section \ref{sec2} below we introduce a spatially asymmetric lattice version of the 3D Ginzburg-Landau model with a
no-monopole constraint, and 
discuss its symmetries.  It is not intended to be a realistic model of high temperature superconductors, but rather to illustrate
certain features which we believe are relevant to the superconductor to pseudogap transition in cuprate 
materials.\footnote{For a recent, quite different approach to an effective gauge Higgs model for the cuprates, based on a a fractionalization of the spin density wave order parameter which results in an emergent non-abelian SU(2) gauge symmetry, see \cite{Sachdev}.}
In section \ref{sec3} we briefly review the center vortex confinement mechanism in SU(N) gauge theories, and the 
importance of global center symmetry in such theories.  The results of lattice Monte Carlo simulations of the modified Ginzburg-Landau model are presented in section \ref{sec4}.  Section \ref{sec5} explores possible connections to
spin glasses and the concept of custodial symmetry breaking, and in section \ref{sec6} we outline how one might derive and study, with our suggested gauge field observables, a more realistic model of the cuprates.
Section \ref{sec7} contains concluding remarks.
    
\section{\label{sec2} The model}

    We begin with a lattice version of the classical Ginzburg-Landau action (i.e.\ no time derivatives), which is also known
as the $D=3$ dimensional abelian Higgs model,with a double charged Higgs field
\bea
     S_{GL} &=& - \b \sum_{x}\sum_{\m=1}^2 \sum_{\n=\m+1}^3 \cos(\th_{\m\n}(x)) \non \\
            & & - \sum_{x} \sum_{\m=1}^3 \text{Re}[ \phi^*(x) e^{2i\th_\m(x)} \phi(x+\hat{\m}) ]  \non \\
           & & + \sum_x  \left[3\phi^*(x)\phi(x) + \l(\phi^*(x)\phi(x)-\g)^2 \right] \ ,
\label{GL}
\eea
where
\beq
\th_{\m\n}(x) = \th_\m(x) + \th_\n(x+\hat{\m}) - \th_\m(x+\hat{\n}) - \th_\n(x) \ .
\label{thmunu}
\eeq
We will simplify further by taking the limit $\l \ra \infty$, and after rescaling the Higgs field and dropping a constant
we have~\footnote{The couplings $\b,\g$ in natural units are $\b=1/(e^2 kT a), ~
\g = g/(kT a)$, where $e$ is electric charge, $g$ is a dimensionless lattice coupling, $T$ is temperature, and $a$ is the lattice spacing.} 
\bea
      S' &=&  - \b \sum_{x}\sum_{\m=1}^2 \sum_{\n=\m+1}^3 \cos(\th_{\m\n}(x))  \non \\
          & & - \g \sum_{x} \sum_{\m=1}^3 \text{Re}[ \phi^*(x) e^{2i\th_\m(x)} \phi(x+\hat{\m})] \ , 
\label{Sp}
\eea
with the unimodular constraint $\phi^*(x) \phi(x) = 1$.    

   The compactness of the U(1) gauge group has one consequence which, in the present context, is very unphysical, namely the existence of magnetic monopoles.  These are lattice artifacts which are responsible, in pure compact U(1) gauge theory, for confinement in $D=3$
spacetime dimensions.  In order to suppress these objects entirely we insert a constraint in the integration measure which prevents their appearance.  The number of monopoles at a site on the dual lattice, in $D=3$ dimensions, is determined from the $\th_\m(x)$ angular variables by the DeGrand-Toussaint \cite{DeGrand:1980eq} construction.  The no-monopole constraint \cite{Mack:1979gb} is a Kronecker delta in the lattice measure which ensures that the monopole number is zero at every site of the dual lattice.

   In cuprates the pairing phenomenon occurs, by some mechanism, in two dimensional planes, while the electromagnetic field extends, as usual, in three space dimensions.  In order to include some remnant of this feature in our model, we simply eliminate the hopping term for the Higgs field in the third spatial dimension
\bea
       S_{MGL} &=&  - \b \sum_{x}\sum_{\m=1}^2 \sum_{\n=\m+1}^3 \cos(\th_{\m\n}(x)) \non \\
          & & - \g \sum_{x} \sum_{\m=1}^2 \text{Re}[ \phi^*(x) e^{2i\th_\m(x)} \phi(x+\hat{\m}) ]  \ ,
\label{MGL}
\eea  
while retaining the unimodular constraint on the Higgs field.   This ``modified Ginzburg-Landau" action, together with the no-monopole constraint, is the theory we will focus on.   It is of course not intended as a realistic effective action for high $T_\text{c}$ phenomena.  The intention is only to illustrate one particular aspect mentioned in the Introduction, namely, the nature of the transition between a Higgs phase, and a vortex liquid (or ``confining'') phase, which we think may have some relevance to the superconducting to pseudogap transition in the cuprates.

  The action $S_{MGL}$ is invariant under three distinct symmetries: 
\begin{enumerate}
\item local U(1) gauge symmetry;
\item global Z$_\text{2}$  symmetry;
\item a set of global U(1) symmetries in the Higgs sector, one for each $xy$ plane.  
\end{enumerate}

\subsection{Gauge Symmetry}

We need not elaborate on local U(1) symmetry, apart from making one important point.  Some textbooks on quantum field theory
erroneously describe the Higgs phase of the theory, which is the phase of superconductivity in the condensed matter context,
as a phase in which the local gauge symmetry is spontaneously broken.  The description is erroneous for the simple reason
that a local gauge symmetry cannot break spontaneously, as proven many years ago by Elitzur \cite{Elitzur:1975im}.  In fact, for a Higgs field with
a single unit of charge, there is no thermodynamic transition in the $\b-\g$ plane which completely isolates the confined and Higgs 
regions of the theory.  The proof is due to Osterwalder and Seiler \cite{Osterwalder:1977pc}, and its implications were elucidated by Fradkin and Shenker \cite{Fradkin:1978dv}.  One consequence, which applies to the double-charged Higgs case as well, is that neither the magnitude nor the phase of the Higgs field $\phi$ can be regarded as an order parameter, since
\begin{itemize}
\item $\langle \phi \rangle = 0$ at all $\b,\g$ in the absence of gauge fixing;
\item $\langle \phi \rangle = 1$ at all $\b,\g$  in unitary gauge, even in the massless phase;
\item in other gauges  $\langle \phi \rangle$ may be zero or non-zero at a particular $\b,\g$, depending on 
the gauge choice \cite{Caudy:2007sf}.
\end{itemize}
This doesn't mean that there is no precise distinction between, say, the Higgs and confinement regions. 
It does mean that a fictitious breaking of the gauge symmetry cannot be used to make that distinction. 

\subsection{\label{z2} Global Z$_\text{2}$  Symmetry}

 In the case of a double-charged Higgs field,  the Higgs phase is distinguished by the spontaneous breaking of a global Z$_\text{2}$  symmetry.  This global transformation can be applied to gauge link variables
\beq
          U_\m(x) = e^{i\th_\m(x)}
\eeq
on any given plane orthogonal to one of the coordinate axes.  

    Consider, e.g., any $y,z$ plane at constant $x$, e.g.\ $x=1$, and make the transformation
\bea
           U_1(\vx) &\ra& \s U_1(\vx) ~, ~~ x=1, ~\mbox{all~~} y,z \non \\
           \s &=& \pm 1 \in Z_2 \ .
\label{Z2trans}
\eea
The action $S_{MGL}$ is invariant under this transformation.  It is also invariant under transformations in any other plane:
\bea
  U_3(\vx) &\ra& \s U_3(\vx) ~, ~~ z=1, ~\mbox{all~~} x,y \non \\
 U_2(\vx) &\ra& \s U_2(\vx) ~, ~~ y=1, ~\mbox{all~~} x,z \ .
\eea
where indices 1,2,3 correspond to spatial directions $x,y,z$ respectively

   A Polyakov line is a Wilson loop along a line running in either of the $x,y,z$ directions, which is 
closed by lattice periodicity; e.g.
\beq
         P(y,z) = \prod_{x=1}^{N_x} U_1(x,y,z) \ ,
\label{Polyakovline}
\eeq
where $N_x$ is the number of lattice sites in the $x$ direction.  Under the Z$_\text{2}$  transformation \rf{Z2trans}, the Polyakov line transforms by $P(y,z) \ra \s P(y,z)$.  We take the lattice extension in the $x$ direction to be arbitrarily large but fixed, and take limit of large extension in the $y,z$ directions. Since the action is invariant under the global Z$_\text{2}$  symmetry, but the Polyakov line is not,  the expectation value $\langle P \rangle$ is, in the limit of large $y,z$ area, a gauge-invariant order parameter for the spontaneous breaking of this symmetry.   The Polyakov line expectation value is often applied, in QCD studies, to detect the high-temperature deconfinement phase.  But it also serves to detect the breaking of global Z$_\text{2}$  symmetry in the Higgs/superconductor phase, and to rigorously distinguish that phase from other phases of the system, when the scalar field carries two units of electric charge.

\subsection{Global $U(1)$ symmetries}

   The action $S_{MGL}$ is also invariant under transformations of the Higgs field
\beq
           \phi(\vx) \ra e^{i\a(z)} \phi(\vx) \ ,
\label{globalU1}
\eeq
which are local in the $z$-direction, but global in any $x-y$ plane; these can be regarded as a set of
independent global U(1) transformations on each $x-y$ plane.   A related symmetry in the Higgs sector, sometimes known as
``custodial symmetry'' (see, e.g., \cite{Willenbrock:2004hu,Maas:2019nso}) does play a role in non-abelian gauge-Higgs theories when the Higgs field is in the fundamental representation, and may even (despite the Fradkin-Shenker  argument \cite{Fradkin:1978dv} based on the Osterwalder-Seiler theorem \cite{Osterwalder:1977pc}) serve to distinguish a Higgs from confinement phase in such theories \cite{Greensite:2018mhh}.  In the present case this global symmetry in the
$x-y$ planes might appear to be irrelevant, owing to the fact (the Mermin-Wagner theorem) that continuous global symmetries cannot break in two dimensions.  Nevertheless, we believe that this symmetry does play a
role in distinguishing the gapped from ungapped phases.  The discussion will be postponed to section \ref{sec5}.

\section{\label{sec3} Center symmetry and center vortices}

    In this section we will take a short excursion into confinement physics in SU(N) non-abelian gauge theories, before returning to the abelian theory described by $S_{MGL}$.  The relevance of center vortices
to confinement was first pointed out by `t Hooft \cite{tHooft:1977nqb}; an extensive review of the confinement mechanisms which have been proposed for non-abelian gauge theories is found in ref.\ \cite{Greensite:2011zz,Greensite:2003bk}.  Here we provide only the briefest summary of ideas which are directly relevant to this article.    
    
    The center of a group is the set of all elements which commute with all other elements of the group.  For an SU(N) group this is
 the set
 \beq
            \{ z_n \mathbbm{1} = e^{2\pi i n/N} \mathbbm{1}, ~ n=0,1,...,N-1 \} \ ,
 \eeq
where $\mathbbm{1}$ the the $N\times N$ unit matrix, and Z$_N \in$ SU(N) is the subgroup composed of these center elements.  The N-ality $k$ of a group representation $R[g], g \in SU(N)$ is defined by the representation of the center subgroup, i.e.
\beq
            R[z g] = z^k R[g] ~~~ \mbox{for~~~} z \in Z_N \ .
\eeq
The fundamental representation has N-ality $k=1$, and the adjoint representation has N-ality $k=0$.   An SU(N) gauge theory with either no matter fields, or with matter fields only in zero N-ality representations, has a global Z$_\text{N}$ center symmetry whose unbroken or broken
realization corresponds to the presence or absence of confinement.  ``Confinement'' means here that the interaction potential between static test charges in the fundamental and anti-fundamental representations, at large color charge separation $R$, rises linearly with $R$ as $R \ra \infty$.

     An example of a global center transformation in an SU(N) lattice gauge theory is a transformation applied to all timelike link variables $U_0(\bx,0)$ at time $t=0$:
\beq
         U_0(\bx,0) \ra z U_0(\bx,0)  ~~~,~~~z \in Z_N \ .
\eeq
It is easy to check that the action, and any contractible Wilson loop, is invariant under this transformation.  On the other hand a Polyakov loop, which is a Wilson loop winding once around the lattice in the periodic time direction, transforms as ${P \ra z P}$.  Since the expectation value of
$P$ is the exponential of minus the free energy of an isolated charge, it follows that color charges are confined 
if $\langle P \rangle = 0$ and center symmetry is unbroken, while they are unconfined in the opposite case $\langle P \rangle \ne 0$ and center symmetry is broken.  

    One of the most striking features of confinement in an SU(N) gauge theory with center symmetry is the fact that the confining force
between color charges, at sufficiently large charge separation, is sensitive only to the N-ality of the color charges, rather than the particular group representation of that N-ality.  In other words, let $W_r(C)$ represent the expectation value of a Wilson loop around closed contour $C$, with the gauge field in representation $r$.  Then for large loops
\beq
           W_r(C) \sim e^{-\s_k A(C)} \ ,
\eeq
where $A(C)$ is the minimal area enclosed by the loop, and $k$ is the N-ality of representation $r$.  The point is that the string tension $\s_k$ depends only on N-ality of $r$.  If we are to attribute confinement to some special class of configurations which dominate the functional integral at large scales, then we must look for configurations which affect loops in different representations, but with the same N-ality, in the same way.  The only known configurations which have this property are called ``center vortices.''

    In a time slice in $D=4$ Euclidean dimensions, a center vortex is a tubelike structure closely analogous to an Abrikosov vortex in superconductivity, in the sense of  being a field configuration carrying a quantized amount of (something analogous to) magnetic flux.  The action density of such configurations is concentrated
in a region of codimension two.  This means that a center vortex is point-like in two Euclidean dimensions, line-like in three dimensions, and surface-like in four dimensions (one may imagine a tube sweeping out a surface-like region in time), with the qualifier ``like'' meaning that in each case the vortex region has a finite thickness.  For an Abrikosov vortex
\bea
             W(C) &\equiv& \exp\left[i{e\over \hbar}\oint_C d\bx \cdot A \right] \non \\
                       &=& -1 \ ,
\eea
where the loop $C$ runs around the vortex, outside the vortex core.  The analogous statement in a non-abelian gauge theory is that if one creates a center vortex topologically linked to a Wilson loop in a representation $r$ of N-ality $k$ running around contour $C$, the loop is transformed by a center element $z \ne 1$, i.e.
\beq
           W_r(C) \ra z^k W_r(C) \ .
\eeq

\subsection{Confinement}

    Confinement in the vortex picture works as follows.  Let the gauge group be SU(2) for simplicity.  Consider a plane of area $L^2$ which is pierced, at random locations, by $N$ center vortices, and consider a Wilson loop of area $A$, in a representation of N-ality $k=1$ lying in that plane.  Then the probability that $n$ of those $N$ vortices will lie inside the area $A$ is
\beq
          P_N(n) = {N \choose n}\left( A\over L^2 \right)^n \left(1 - {A \over L^2} \right)^{N-n} \ .
\eeq
Each vortex piercing the Wilson loop contributes a factor of $-1$, so the vortex contribution to the Wilson loop is
\beq
     W(C) = \sum_{n=0}^N (-1)^n P_N(n) = \left(1- {2A \over L^2} \right)^N \ .
\eeq
Now keeping the vortex density $\rho=N/L^2$ fixed, and taking the $N,L \ra \infty$ limit, we arrive at the Wilson loop area law falloff
\beq
W(C) = \lim_{N\ra \infty} \left(1 - {2\rho A \over N} \right)^N = e^{-2 \rho A} \ .
\eeq
That is the center vortex confinement mechanism in three lines \cite{Engelhardt:1998wu}.  
It is the simplest such mechanism known.  The crucial assumption is that vortex piercings in the plane are random and uncorrelated, and this implies that vortices percolate throughout the spacetime volume.  

    There is a great deal of numerical evidence in favor of this picture, obtained from lattice Monte Carlo simulations.  Most of this numerical work makes use of a technique, known as ``center projection,'' for locating center vortices in lattice configurations.  The idea is to map SU(N) lattice configurations into $Z_N$ configurations, which have only vortex excitations.  This is accomplished by a gauge transformation into ``maximal center gauge,'' which brings the SU(N) link variables as close as possible, on average, to the $Z_N$ center elements of the group.  Maximal center gauge maximizes the quantity
\beq
          R = \sum_{x,\m} |\tr[U_\m(x)|^2 \ ,
\eeq
which is equivalent to Landau gauge fixing of link variables in the adjoint representation.  One then maps each link variable to the closest $Z_N$ center element.  What is remarkable is that the center projected configurations are qualitatively, and to a large extent quantitatively, similar to the full SU(N) configurations, in terms of confinement, chiral symmetry breaking, and even the mass spectrum.  There is also a simple technique for removing center vortices from the SU(N) configurations.  When this is done, confinement and chiral symmetry breaking disappear.  For older reviews, see \cite{Greensite:2011zz,Greensite:2003bk}.  For more recent developments, see \cite{Kamleh:2017thj,Trewartha:2017ive}.

\subsection{The Higgs phase}

   Confinement is lost, in a non-abelian theory in $D \le 4$ dimensions, when the global center symmetry of the action is broken spontaneously, either at high temperatures (this is known as the ``deconfinement'' transition), or via a transition to a Higgs phase.
In the latter case, the action contains one or more Higgs fields $\phi$ transforming in the adjoint representation of the gauge group. On the lattice, in $d$ Euclidean spacetime dimensions, the Higgs action has the form
\bea
     S_H &=& - \sum_x \sum_{\m=0}^{d-1}  \text{Re}[\phi^\dg(x) U^A_\m(x) \phi(x+\hat{\m}) \non \\
              & & + \sum_x \{d \phi^\dg(x) \phi(x)  + V[\phi(x)] \}  \ ,
\eea
where $V(\phi)$ is the Higgs potential, and the superscript $A$ in $U^A_\m$ means that the link variables are taken to be
in the adjoint representation of SU(N).  Since $U^A$ is invariant under transformations $U \ra zU$, where $z \in Z_N$, the gauge-Higgs action is invariant under the global center symmetry defined above. 

    The distinction between the confinement and Higgs phases of gauge theories with adjoint Higgs fields is nicely represented by the behavior of the Wilson loop $W(C)$ and its dual in $D=4$ dimensions, known as the `t Hooft loop  $B(C)$ 
\cite{tHooft:1977nqb}, which can be thought of as a center vortex creation operator. In the confinement phase, Wilson loops fall with the area and the expectation value of `t Hooft loops fall with the
perimeter of the loop; in the Higgs phase it is the reverse. 

     Alternatively, on a finite lattice the Polyakov line is defined by  \rf{Polyakovline},
only generalized to the non-abelian gauge group and (on the lattice) the SU(N) link variables.
Now if we take  one of the Euclidean directions (say $\m=0$) to be the time direction, the Polyakov line in the
time direction is
\beq
         P(\vx) = \tr\left[ \prod_{t=1}^{N_t} U_0(\vx,t) \right]  \ .
\label{Polyakov_nab}
\eeq
This observable is gauge-invariant, but transforms by a center element $z \in Z_N$ under a global center transformation.
It is therefore an order parameter for spontaneous symmetry breaking of global center symmetry.  If we keep the
time extension arbitrarily large but fixed, and take the large volume limit in the remaining space directions, then the Higgs
phase is the phase in which $\langle P \rangle \ne 0$.  This is because the Polyakov line is related to the
free energy $F_q$ of an isolated static color charge by
\beq
           \langle P \rangle = e^{-N_t F_q} \ .
\eeq
It follows that when $\langle P \rangle = 0$ the free energy of an isolated charge is infinite, and quarks are confined.  Conversely,
when $\langle P \rangle \ne 0$ the free energy is finite, and quarks are unconfined.  In this sense, keeping one (time) direction
constant, although arbitrarily large, in the limit that the lattice extension in the space directions are taken to infinity, we may
say that the Higgs phase is a phase of spontaneously broken center symmetry.

     The analogy we pursue in this paper is that the pseudogap phase in the cuprates is, in the same sense, a phase of unbroken Z$_\text{2}$  global symmetry, and corresponds to the confinement phase
in an SU(N) gauge theory, which is a phase of unbroken $Z_N$ center symmetry. These phases can each be regarded
as a vortex liquid of some kind.  Likewise, the superconducting phase in the cuprates, and the Higgs phase in a gauge theory,
correspond to the spontaneously broken phase of global Z$_\text{2}$  and $Z_N$ symmetry, respectively.

     In the case of SU(N) gauge theories such as QCD, with matter in the fundamental representation of the gauge group,
the action breaks global center symmetry explicitly, $\langle P \rangle$ is always non-zero, and Wilson loops fall off asymptotically
with a perimeter law.  Moreover there is no thermodynamic transition isolating the Higgs from the confinement regions \cite{Osterwalder:1977pc,Fradkin:1978dv}. One may ask in what sense these theories are confining, apart from the fact that the asymptotic spectrum consists of massive color singlets.  This is, in fact, a surprisingly subtle question.  Our view
is presented in ref.\ \cite{Greensite:2017ajx}.

\section{\label{sec4} Numerical Results}

\subsection{Pure gauge field}

    We first consider the three dimensional gauge theory with $\g=0$; i.e.\ no coupling to the scalar field. Our proposed effective theory eliminates monopoles by a constraint.  Without this constraint there is confinement in
2+1 dimensional compact U(1) gauge theory, in the sense of a linearly rising potential between static charges, as we know from the classic work of Polyakov \cite{Polyakov:1976fu}.  With the no-monopole constraint this linear confinement property ought to disappear, and the potential between static charges should increase only logarithmically, as in the free continuum theory.  This is the first thing to check.

\begin{figure}[t!]
\centerline{\includegraphics[scale=0.7]{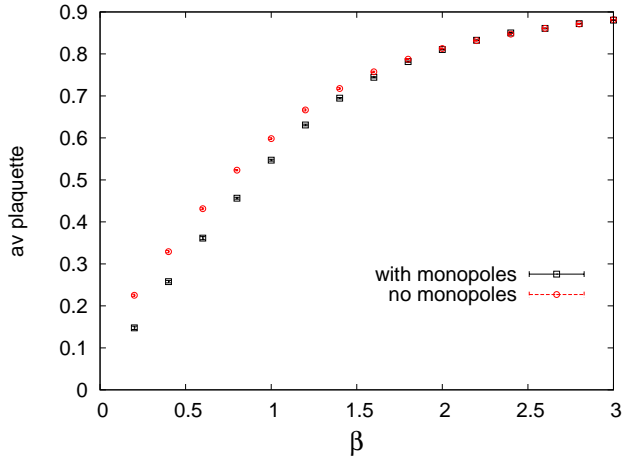}}
\caption{Average plaquette values in pure compact U(1) lattice gauge theory in D=3 dimensions, with and without the no-monopole
constraint.} 
\label{bplaq}
\end{figure} 
    
    Figure \ref{bplaq} is a comparison of the average plaquette $\langle \cos \th_{\m\n} \rangle$ vs.\ $\b$, in compact U(1) theory
with and without monopoles.  The plaquette averages in the two theories converge as $\b$ increases as expected, since the monopole density in the unconstrained theory falls rapidly beyond $\b=1$.  However, there is a finite monopole density at any $\b$, and even if the difference in average plaquette with and without monopoles is negligible, the unconstrained theory has a linear static potential, while the constrained theory does not.  To see this numerically, we note that the potential $V(R)$ between static opposite charges is given by the
logarithmic time derivative of rectangular Wilson loops
\beq
            V(R) = - \lim_{T \ra \infty} {d\over dT} \log W(R,T) \ .
\eeq
On the lattice we extract $V(R)$ from a best linear fit to the data for $-\log W(R,T)$ vs $T$, at $T>10$.  

    Of course the word  ``potential'' should not be taken too literally in this particular context.  $V(R)$ is indeed the potential between static charges in a U(1) theory with two space dimensions and one time dimension, and a linearly rising potential
would imply confinement of electric charge. But in three space dimensions it is simply
a diagnostic of the behavior of $W(R,T)$ at large $R$ or large $T$ due to thermal fluctuations of the magnetic $\bf B$ field.  
In particular, if $V(R)$ is asymptotically linear at large $R$, this just means that the Wilson loop falls off exponentially with the area $RT$ enclosed by the loop.  
     
\begin{figure}[t!]
\centerline{\includegraphics[scale=0.7]{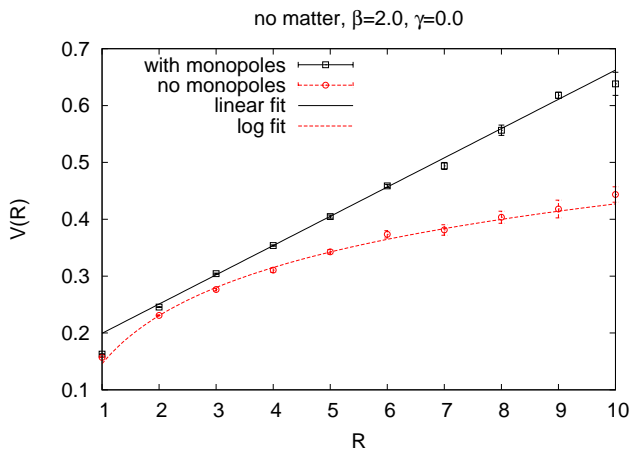}}
\caption{The potential between static charges of opposite sign in pure compact U(1) gauge theory, D=3 dimensions, at gauge coupling
$\b=2.0$, with and without the no-monopole constraint.  The potential rises linearly in the unconstrained case, but only logarithmically with no monopoles, as in the free (non-compact) gauge theory.} 
\label{g0pot}
\end{figure} 

   With that caveat, let us compare the potential $V(R)$ in the compact U(1) theory with and without the no-monopole
constraint.  In a free theory in 2+1 dimensions we expect the potential between static opposite charges to rise logarithmically with charge separation, while in the compact U(1) theory, without  any constraint, one expects to see a linear potential.
The result of a simulation at  $\b=2$ is shown in Fig.\ \ref{g0pot}.  We find that the potential rises logarithmically in the compact U(1) theory, as in the free theory, when a no-monopole constraint is imposed.  The unconstrained compact U(1) theory displays a linearly rising potential, as expected. Note that this drastic difference in the potential is displayed at a coupling $\b=2$ where we 
also see, from Fig.\ \ref{bplaq}, that the difference in average plaquette values in the constrained and unconstrained theories
is almost imperceptible.
 
\subsection{Modified Ginzburg-Landau}

    We now couple the scalar field to the gauge field by setting $\g > 0$ in \rf{MGL}.  The first task is to
determine the phase diagram in the space of $\b-\g$ couplings.  There is only one symmetry which can be spontaneously broken, namely the global Z$_\text{2}$  symmetry discussed in section \ref{z2}, and the appropriate order parameter is a Polyakov line
running in a direction parallel to the $x$ or $y$ axes.  The superconducting region can only be the region where this global Z$_\text{2}$  symmetry is broken, and to check this we look for evidence in the potential, extracted from Wilson loops, that the photon has acquired a mass.

\begin{figure}[t!]
\subfigure[~Polyakov line]  
{   
 \label{poly}
 \includegraphics[scale=0.55]{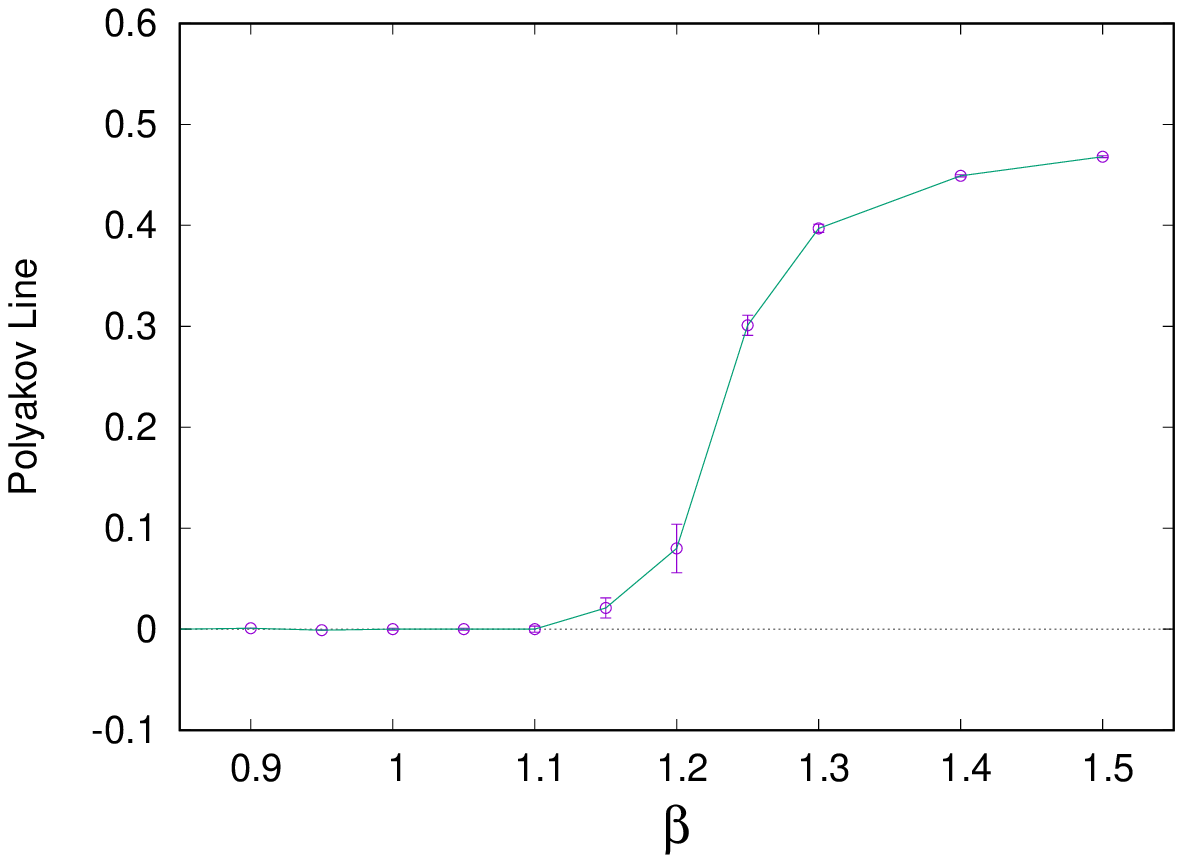}
}
\subfigure[~Susceptibility]  
{   
 \label{psus}
 \includegraphics[scale=0.55]{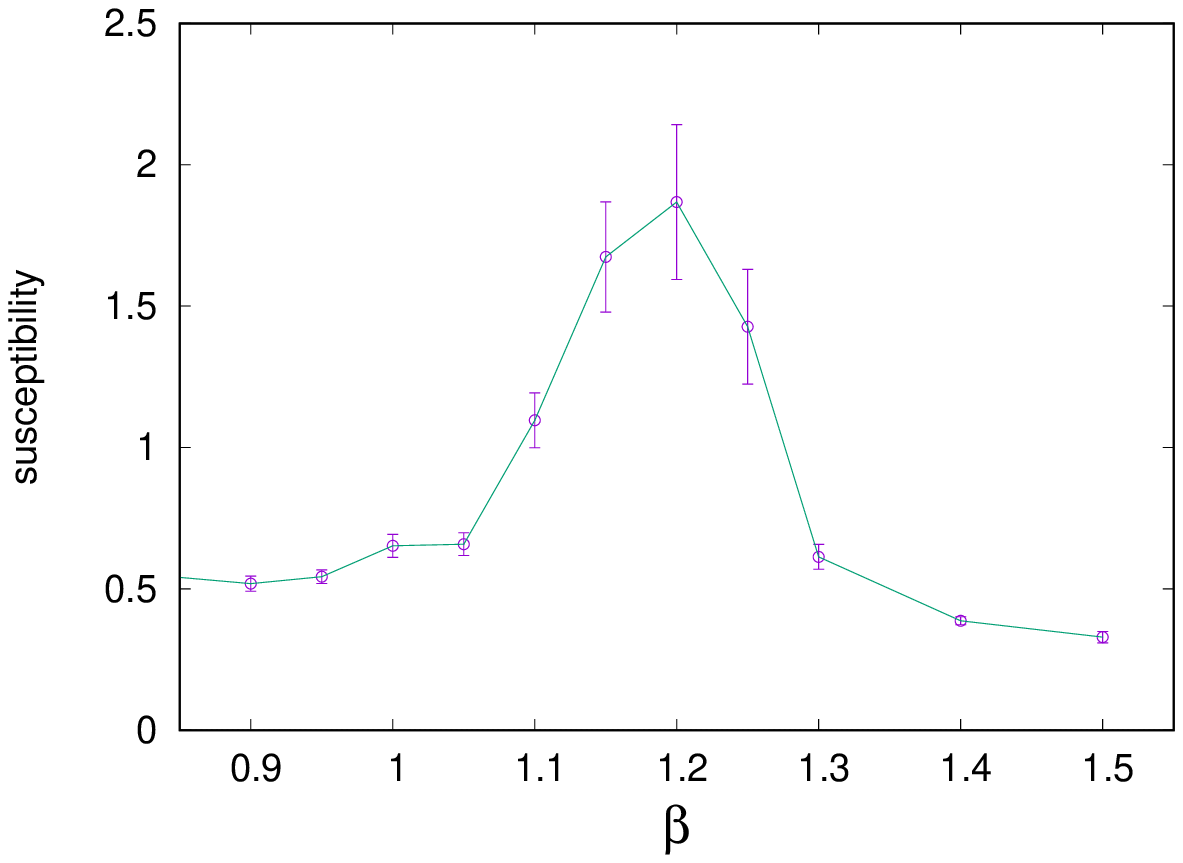}
}
\caption{(a) Polyakov lines and (b) Polyakov line susceptibilities vs.\ $\b$ at $\g=6$ on a $40^3$ lattice volume.  All Polyakov
lines are 40 lattice units in length, computed in the $x$ and $y$ directions in $xy$ planes.}
\label{Polyakov}
\end{figure}

\begin{figure}[t!]
\centerline{\includegraphics[scale=0.7]{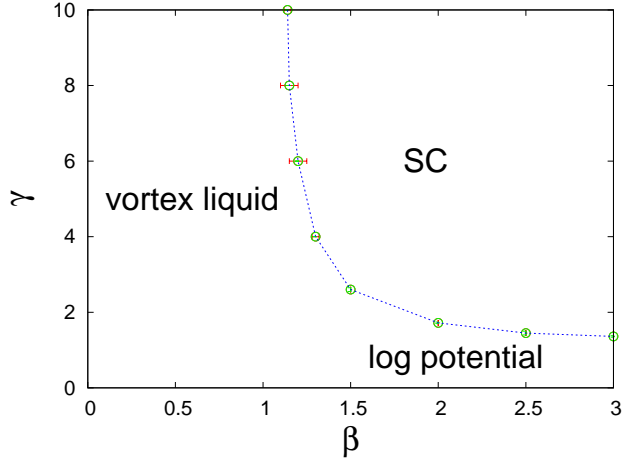}}
\caption{Phase diagram of the modified Ginzburg-Landau theory in the $\b-\g$ coupling plane.  The superconducting
phase is also a phase of broken global Z$_\text{2}$  symmetry.}
\label{bg}
\end{figure}

   Our numerical simulations are carried out on a $40^3$ lattice volume.  In the superconducting region we find $\langle P \rangle
\ne 0$, while in the normal region we have $\langle P \rangle = 0$ within error bars.  The transition points are estimated, on the
cubic lattice, by looking for a peak in the Polyakov line susceptibility, either at fixed $\g$ and varying $\b$, or at fixed $\b$ while varying $\g$.  Examples of our data for the Polyakov line and the Polyakov line susceptibility vs.\ $\b$,
at fixed $\g=6$, are shown in Figs.\ \ref{poly} and \ref{psus} respectively.   The resulting phase diagram is shown in Fig.\ \ref{bg}, but since the phase boundary (just drawn as straight lines between
the numerically determined transition points) is determined from the breaking of global Z$_\text{2}$  symmetry, the labeling of the
different regions (SC, vortex liquid, log potential) must be justified.

\begin{figure}[t!]
\centerline{\includegraphics[scale=0.7]{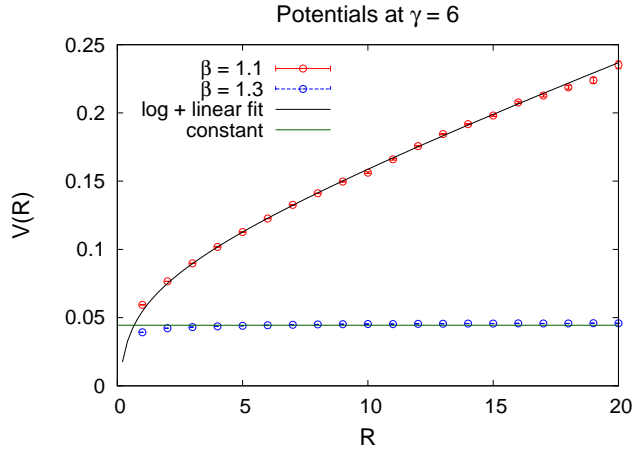}}
\caption{The potential $V(R)$ at $\g=6$ just outside ($\b=1.1$) and just inside ($\b=1.3$) the superconducting region.}
\label{g6pot}
\end{figure} 

  We begin at $\g=6$, comparing $V(R)$ calculated from $W(R,T)$ in the $x-y$ plane, as explained above.  Fig.\ \ref{g6pot}
displays $V(R)$ calculated just outside the SC region, at $\b=1.1$ in a region labeled ``vortex liquid'', and $V(R)$ just inside
the SC region, at $\b=1.3$.  The potential in the vortex liquid region is fit to the form
\beq
         V(R) = a + b \log(R) + \s R \ ,
\label{fitfun}
\eeq
and we find from the fit that $\s = .00623(8)$, i.e.\ a ``confining'' potential, meaning that Wilson loops fall off asymptotically
with loop area.  In contrast, inside the SC region, we see that $V(R)$ is nearly constant for $R>3$, consistent with what one would expect from a finite-range interaction mediated by a massive photon.  This is evidence of superconductivity in the SC region.

   We have labeled the region at small $\b$ and large $\g$, outside the SC domain, as a ``vortex liquid,'' and this characterization
must now be justified.  Let us consider going to unitary gauge, $\phi(x)=1$, and taking the $\g=\infty$ limit.  In this case,
the $U_k(x)$ link variables in the $x,y$ directions are forced to be $U_k = \pm 1$, i.e.\ the variables of a Z$_\text{2}$  gauge theory,
at least in the $xy$ planes.  The only excitations in these planes are at plaquettes where $\cos\theta_{12}(x) =  -1$, and these are Z$_\text{2}$  vortex configurations.  Consider a Wilson loop
\beq
           W(R,T) = \langle U(R,T) \rangle \ .
\eeq
where $U(R,T)$ is a product of U(1) link variables around a rectangle oriented in one of the $x,y$ planes.  Suppose, in some gauge field configuration, there are $n$ plaquettes within this rectangle with $\cos\theta_{12}(x) = -1$.  Then
\beq
           U(R,T) = (-1)^n
\eeq
in this $\g=\infty$ limit.  If there is a finite density of vortices in the plane, and if vortex positions are entirely uncorrelated, then
this leads to an area law falloff of $W(R,T)$, and a linear potential for $V(R)$, as explained in section \ref{sec3}.  If there is some finite range correlation among the vortices, then there will be a deviation from the linear potential up to that finite range.  
A linear potential is therefore the signature that the system in a plane is a disordered gas or liquid of Z$_\text{2}$  vortices.\footnote{Of course it must be kept in mind that even in the $\g=\infty$ limit we are not dealing with a trivial Z$_\text{2}$  gauge theory in two dimensions, since even in this limit the gauge field extends into all three spatial dimensions.  For the action $S'$ in  \rf{Sp}, the 
$\g=\infty$ theory would be Z$_\text{2}$  gauge theory in three dimensions.}
 
    We can try to locate vortices in $xy$ planes away from the $\g=\infty$ limit, with the strategy of (i) performing a gauge transformation which brings link variables in the $xy$ planes as close as possible to $\pm 1$; and (ii) ``Z$_\text{2}$  projection'' 
in the $xy$ planes, i.e. projecting link variables in the $x,y$ directions onto the closest element of the Z$_\text{2}$  subgroup of the U(1) gauge group.   The gauge transformation should maximize the quantity
\beq
           Q = \sum_{x} \sum_{i=1}^2 \cos^2 \th_i(x) \ ,
\eeq
and this is done by performing a sequence of gauge-fixing sweeps of the lattice.  In this gauge there is a remnant local Z$_\text{2}$  gauge symmetry.  Gauge transformations are made site-by-site,
at each site making a transforming which maximizes
\beq
               \sum_{i=1}^2 \Bigl[\cos^2 \th_i(x) + \cos^2 \th_i(x - \hat{i}) \Bigr] \ .
\eeq
This procedure converges to a local maximum of $Q$.\footnote{Finding the global maximum is likely to be an NP hard problem.}  The gauge fixing sweeps end when the fractional increase in $Q$ from one sweep to the next falls below $10^{-8}$.  Z$_\text{2}$  projection
consists of the mapping
\beq
          U_i(x) \ra Z_i(x) = \text{sign}[\text{Re}(U_i(x))]  ~~,~~ i=1,2 \ .
\eeq

\begin{figure}[t!]
\centerline{\includegraphics[scale=0.7]{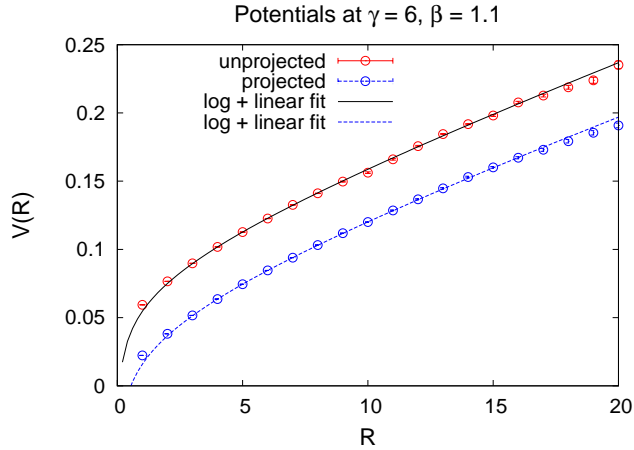}}
\caption{A comparison of the projected and unprojected potentials, $V(R)$ and $V_{proj}(R)$, at a point
in the vortex liquid region.}
\label{wz611}
\end{figure}

    We define $Z(R,T)$ as the product of projected link variables $Z_i(x)$ around an $R\times T$ rectangle, with the corresponding expectation values
 \beq
           W_{proj}(R,T) = \langle Z(R,T) \rangle \ ,
\eeq
and we compute the projected potential $V_{proj}(R)$ from the $W_{proj}(R,T)$ by the same procedure used to obtain $V(R)$ from $W(R,T)$.   In Fig.\ \ref{wz611} we compare $V(R)$ and $V_{proj}(R)$ vs.\ $R$, at $\g=6, ~\b=1.1$, and it can be seen that
the projected and unprojected potentials are essentially parallel, differing only by a constant $R$-independent self-energy.
Since the vortices alone, in the projected configuration, reproduce the potential in the unprojected lattice, it seems reasonable to attribute the $R$-dependence of the potential to the effects of vortices, whose positions in the unprojected lattice are located by
the excitations in the projected lattice.   As a further check
we can compute the average value of $\cos\th_{\m\n}(x)$ for plaquettes on the original lattice, at locations where the plaquette
on the projected lattice is $-1$, indicating the presence of a Z$_\text{2}$  vortex.  At couplings $\b=1.1, \g=6$, these special ``vortex plaquettes'' have an average value of $0.398$, to be compared with the average over all plaquettes, which is 
$\langle\cos(\th_{\m\n}(x)\rangle=0.909$.  So although the procedure for locating vortices involves fixing to a particular gauge (i.e.\ maximal Z$_\text{2}$  gauge), we nevertheless find that the locations of vortices on the projected lattice are very strongly correlated with a gauge-invariant observable on the unprojected lattice, i.e.\ the gauge-invariant field strength.\footnote{The fact that plaquettes on the unprojected lattice, at vortex locations on the projected lattice, are not closer to $-1$ can be attributed to either a thickness of
the vortex which is greater than one lattice spacing, and/or a small error, on the projected lattice, in finding the actual vortex
location.}

\begin{figure}[t!]
\centerline{\includegraphics[scale=0.7]{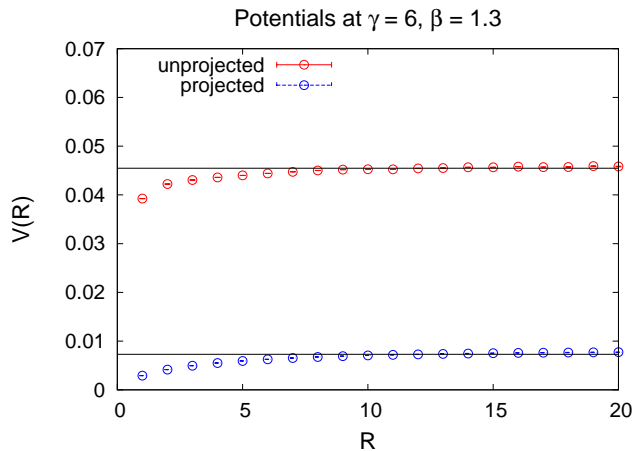}}
\caption{The projected and unprojected potentials at a point inside the SC phase.}
\label{wz613}
\end{figure}

   For these reasons we label the region where a linear potential can be identified, and where the projected and unprojected string tensions agree, as a ``vortex liquid.'' The linear potential disappears in both the projected and unprojected potentials inside the SC region, as seen in Fig.\ \ref{wz613} for $\g=6,~\b=1.3$.

\begin{figure}[t!]
\centerline{\includegraphics[scale=0.7]{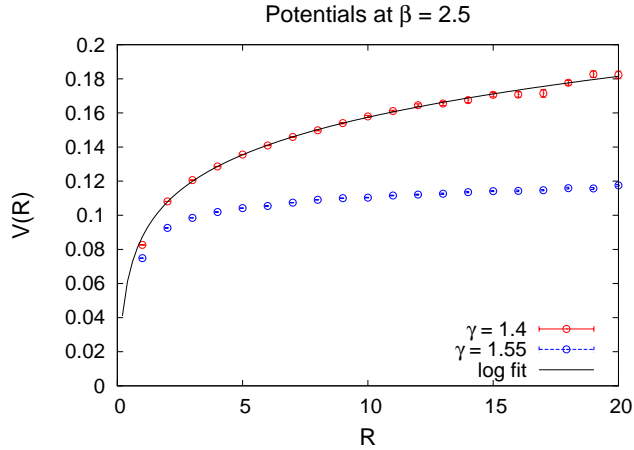}}
\caption{The potential $V(R)$ at $\b=2.5$ just below ($\g=1.4$) and just above ($\g=1.55$) the superconducting transition line. }
\label{wsB}
\end{figure}

    Linear confinement cannot persist down to $\g=0$, simply because the theory in that limit is pure U(1) gauge theory,
and with the no-monopole constraint there are no topologically stable configurations which could disorder Wilson loops.
And at small but finite $\g$ we cannot, in fact, detect any string tension from numerical simulations.  Fig.\ \ref{wsB}
shows our data for $V(R)$ at $\b=2.5$ just below ($\g=1.4$) and just inside ($\g=1.55$) the SC phase.  Just below the SC
phase, at $\b=2.5,~\g=1.4$, the potential fits a logarithm, i.e.\ it is consistent with eq.\ \rf{fitfun} with $\s \approx 0$, which is the fit shown in Fig.\ \ref{wsB}.  Inside the SC phase at  $\b=2.5,~\g=1.55$ the potential is nearly flat, as expected.

\begin{figure}[t!]
\centerline{\includegraphics[scale=0.7]{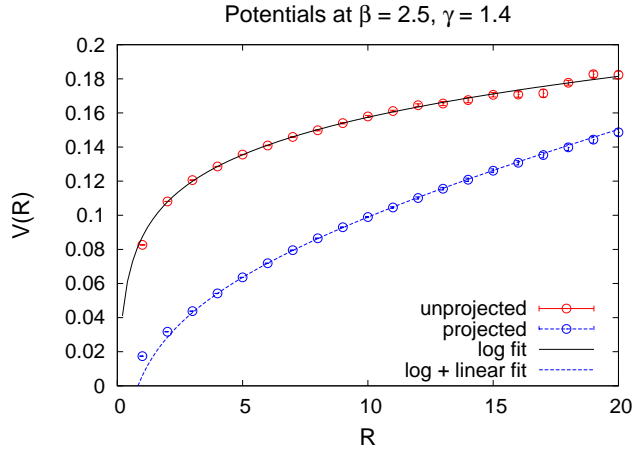}}
\caption{A comparison of the projected and unprojected potentials in the normal phase, at $\b=2.5,~\g=1.4$.  In this
case the Z$_\text{2}$  projection is misleading.}
\label{wz2514}
\end{figure}

   In a region where the potential is logarithmic, i.e.\ essentially perturbative, we would not expect to explain the potential
via purely non-perturbative effects due to vortices.  In this region Z$_\text{2}$  projection should fail to match the unprojected
potential, and in fact that is what we see in Fig.\ \ref{wz2514}, where the projected and unprojected potentials are compared
at $\b=2.5,~\g=1.4$.  The unprojected potential fits \rf{fitfun} with $\s \approx 0$, as already noted.  Not so for the projected
potential, where we find $\s=0.00322(5)$.  Morever, the average plaquette value in this case is $0.887$, while the average
value of plaquettes whose location coincides with vortices on the projected lattice is $0.803$.  While there is some modest correlation here between vortex location on the projected lattice and plaquette value on the unprojected lattice, it is greatly reduced as compared (0.909 vs.\ 0.398) to the previous case at $\b=1.1, \g=6$, in a region described as a vortex liquid.
   
    So the normal phase appears to have regions with and without a string tension, associated with the presence or absence, respectively, of vortex effects.  We have been unable, however, to detect a thermodynamic phase transition between the vortex liquid and  logarithmic potential regions, and it is numerically somewhat challenging to pin down exactly where the string tension disappears.     To search for a thermodynamic transition from the behavior of the plaquette susceptibility we have scanned the phase diagram at fixed $\b=1.1$ and $0 < \g < 6$, and also at fixed $\g=1.4$ and $0<\b < 2.5$.  We have not found any
evidence of a transition along these search lines, which cross from the vortex liquid to the log potential regions. The absence of a thermodynamic phase transition is perhaps unsurprising, since there is no symmetry which distinguishes the vortex liquid from the log potential regions. Indeed the vortex liquid and log potential regions are both ``confining'' in the sense that 
$V(R) \ra \infty$ as $R \ra \infty$ in each case, and consequently $\langle P \rangle= 0$ in both regions.

\begin{figure}[t!]
\centerline{\includegraphics[scale=0.7]{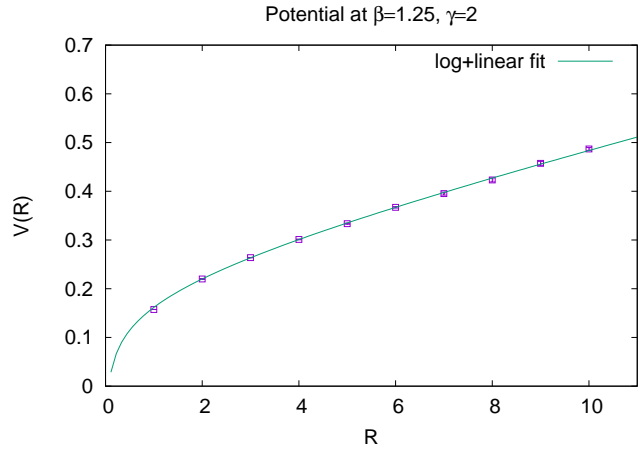}}
\caption{The potential at $\b=1.25,\g=2.0$ on a $40^3$ lattice volume, computed at $R \le 10$.}
\label{b125g2}
\end{figure} 
    
       If it were possible to compute $V(R)$ out to $R=20$ or larger everywhere in the phase diagram, then it might be possible to pinpoint the disappearance of the linear potential, but this strategy is frustrated, at small $\b,\g$, by
very large error bars on relatively small Wilson loops.  As $\g$ is reduced at small $\b$, large Wilson loops become noisy, and we are not able to measure $V(R)$ up to the limit set by the lattice size.   As an example, we show in Fig.\ \ref{b125g2} our results at 
$\b=1.25,~\g=2$, together with a best fit to \rf{fitfun}.  In this case we still find evidence of a linear potential with $\s=0.0226(6)$; a purely logarithmic fit fails completely.  However, at these couplings we cannot reliably go beyond $R=10$ on the $40^3$ lattice volume.  Moreover, as $\g$ is reduced the Z$_\text{2}$  projection becomes increasingly inaccurate, e.g.\ at  $\b=1.25,~\g=2$ the
Z$_\text{2}$  projected string tension is about 30\% larger than the the string tension derived from the unprojected data.  As we increase $\b$ at fixed $\g=1.4$ it is possible to again measure the potential at larger values of $R$, but the string tension seems to either gradually disappear, or else exists at $R$ values beyond the practical limitations imposed by statistics and lattice size.  In any case we cannot detect any trace of a linear potential at $\b=2.5,~\g=1.4$, as seen in Fig.\ \ref{wz2514}.

The precise manner in which the disordering effects of the vortex liquid disappear in the normal phase as $\g$ is reduced and $\b$ increased, whether that disappearance is sudden or gradual, and whether it is associated with some instability in the vortex configurations at small $\g$, is unclear at the moment.  In the next section we will suggest that the transition
from the logarithmic region to the linear potential region may be related to a spin glass transition.

\section{\label{sec5} A spin glass phase?}

    Although Elitzur's theorem rules out the breaking of a local gauge symmetry, it is always possible to impose a gauge
condition, such as Coulomb or Landau or a maximal axial gauge, which preserves a global subgroup of the gauge group.
Spontaneous breaking of such remnant symmetries is not forbidden by the Elitzur theorem, and some texts 
do define spontaneously broken gauge symmetry in this way, e.g.\ \cite{duncan2012}.\footnote{In this connection, see
also \cite{GREITER2005217}.}  There are at least
two problems with that idea, however.  The first is that transition lines for remnant symmetry breaking may differ in different gauges \cite{Caudy:2007sf}.   The second is that in the cuprates, the Mermin-Wagner theorem forbids spontaneous breaking
of any continuous symmetry, yet cuprates have a superconducting phase, which therefore defies characterization in terms
of the breaking of a continuous symmetry.


   Leaving aside the second issue for a moment, some authors have introduced gauge invariant order parameters for
spontaneously broken gauge symmetry.  It is not hard to construct such order parameters, but all of the ones we
are aware of are based, either explicitly or implicitly, on a gauge choice. Let $g(x;U)$ be a gauge transformation to
a gauge $G$ such as Landau or Coulomb or maximal axial gauge, and we consider the order parameter 
\beq
          Q_x = g(x;U) \phi(x)
\eeq    
where $\phi(x)$ is the Higgs field.  By construction, $Q_x$ is invariant under local gauge transformations of $U_\m$ and 
$\phi$, but does
transform under the remnant global gauge symmetry.  Constructions of that type are found in the literature, e.g.\ in \cite{schakel1998boulevard} and \cite{dutch}, where $G$ is (implicitly) an axial gauge, or in \cite{Kennedy:1986ut},
where $G$ is lattice Landau gauge, or the Dirac order parameter (see, e.g., \cite{Hansson:2004wca}) where $G$ is Coulomb gauge.  Although these order parameters are described as (and in fact are) locally gauge-invariant, it should be understood that a certain gauge choice, and therefore a certain arbitrariness, underlies these constructions.  Evaluation of such $Q$ observables, in the absence of gauge fixing, is completely equivalent to evaluating
the expectation value $\langle \phi \rangle$ in a particular gauge, and the case $\langle \phi \rangle \ne 0$ means that
the remnant global symmetry has been broken in that gauge. 
    
    In \cite{Greensite:2018mhh} we have proposed a different identification of the Higgs phase of a gauge-Higgs theory:  the
Higgs phase is the phase of a spontaneously broken custodial symmetry.  The term ``custodial symmetry'' is adopted from the electroweak theory of particle physics, and refers to a global symmetry of the Higgs field which does not transform the
gauge field.   Let us first consider the lattice abelian Higgs action $S_{GL}$ of \rf{GL}, where 
$S_{GL}=S_W(U) + S_m(U,\phi)$ is the sum of a pure gauge Wilson action $S_W(U)$ and the part of the action involving the matter field $S_m(U,\phi)$. In this theory the custodial symmetry is the group of global U(1) transformations 
$\phi(x) \ra e^{i\a} \phi(x)$.  Because the Higgs field transforms also under local gauge transformations, its expectation value vanishes in the absence of gauge fixing, so the question is how to observe the breaking of custodial symmetry without fixing the gauge in some way.  The proposal in \cite{Greensite:2018mhh} was made in the context of a non-abelian gauge Higgs theory, but applies equally to the abelian theory under consideration.  The idea is to write the usual partition
function as a sum of the partition functions of a spin system in an external gauge field, i.e.    
\beq
               Z  =  \int DU ~ Z_{spin}[U] e^{-S_W(U)} \ ,
\label{Z_1}
\eeq
where
\bea
               Z_{spin}[U] &=& \int D\phi ~ e^{-S_m(U,\phi)} 
\eea
Since the gauge field is held constant in $Z_{spin}[U]$, the only symmetry of the ``spin system'' is the group of global transformations $\phi(x) \ra e^{i\a} \phi(x)$, and this symmetry {\it can} break spontaneously, depending on the background gauge field, and the value of $\g$.  For an operator $\Omega(\phi,U)$ we define the expectation value in the spin system
\beq
       \overline{\Omega}(U) = {1\over Z_{spin}(U)} \int D\phi \ \Omega(\phi,U) e^{-S_m(U,\phi)} \ ,
\eeq   
Then the full expectation value is
\bea
          \langle \Omega \rangle &=& {1\over Z} \int DU D\phi \ \Omega(\phi,U) e^{-S_{GL}} \non \\
                                   &=& \int DU \  \overline{\Omega}(U)  P(U)
\eea     
where the spin system average $\overline{\Omega}(U)$ is evaluated in a background with $U$ chosen from the probability distribution 
\beq
       P(U) = {1\over Z} Z_{spin}[U] e^{-S_W(U)} \ .
\label{PU}
\eeq
If $\Omega$ is simply the scalar field $\phi(x)$ we may define
\beq
       \overline{\phi}(x;U) = {1\over Z_{spin}(U)} \int D\phi \ \phi(x;U) e^{-S_m} \ ,
\eeq
Because $U_i(x)=\exp[i\th_i(x)]$ is not gauge fixed, and $\th_i(x)$ will in general vary wildly with position,
the same will be true of $\overline{\phi}(x;U)$, and we must expect the spatial average of this quantity to
vanish, even if custodial symmetry in $Z_{spin}(U)$ is spontaneously broken.  The proposal is instead
to take the spatial average of a gauge-invariant and positive definite quantity
\bea
          \Phi  &=& {1\over V} \sum_{x,t} \langle | \overline{\phi}(x;U) | \rangle \non \\
                  &=& {1\over V} \sum_{x,t} \int DU | \overline{\phi}(x;U) | P(U) \non \\
          &=& {1\over Z} \int DU  {1\over V} \sum_{x,t} e^{-S_W(U)} \left| \int D\phi \ \phi(x) e^{-S_m(U,\phi)} \right|
\eea   
Custodial symmetry is said to be spontaneously broken if $\Phi > 0$ in the $V \ra \infty$ limit, but this does
not imply long range correlations in any gauge-invariant observable \cite{Greensite:2018mhh}.  It should
be emphasized that there is no appeal whatever, in this formulation, to any choice of gauge.  

    The order parameter $\Phi$ is very closely related to the Edwards-Anderson \cite{Edward_Anderson} order parameter for
spin glasses.  Their orginal model, for Ising spins $S_x = \pm 1$ and in the absence of an external magnetic field, was
\beq
            H_{EA} = - \sum_{<xy>} J_{xy} S_x S_y 
\eeq
where $J_{xy}$ is a set of random couplings taken from some probability distribution $P(J)$.  Since the couplings are
random, the spatial average of spins will tend to average to zero, so the order parameter is taken to be
\bea
           \kappa &=&   \int \prod_{<ij>} dJ_{ij}  \left( {1\over V} \sum_k (\overline{S}_k)^2  \right) P(J) \non \\
           \overline{S}_i  &=& {\prod_n\sum_{S_n} S_i e^{-H_{EA}} \over \prod_n \sum_{S_n} e^{-H_{EA}}}
\eea            
The analogy between $\Phi$ and $\k$ is obvious.  Instead of spins $S_i$ we have a unimodular complex
field $\phi(x)$, with squared link variables $U^2_j(x)=e^{2i\th_j(x)}$ in $S_m$ which play the role of the random couplings $J_{ij}$, and these
link variables are drawn from the probability distribution $P(U)$ in \rf{PU}, rather than the simpler (usually Gaussian)
distributions assigned for $P(J)$.  The symmetry of the Edwards-Anderson model is global $Z_2$, while in our case
it is global U(1).  And while Edwards-Anderson order parameter  $\k$ involves the square 
$(\overline{S}_i)^2$ of the spin average, our order parameter $\Phi$ involves the modulus $ | \overline{\phi}(x;U) | $.
But the general idea is the same.  

    In  \cite{Greensite:2018mhh}  we have argued, in the context of the SU(2) gauge-Higgs model with the Higgs
field in the fundamental representation of the gauge group, that the Higgs and confinement regions are separated
by the spontaneous breaking of a custodial symmetry, and also that the Higgs and confinement regions are distinguished
physically by different realizations of confinement, which we have termed  color (C) confinement in the Higgs region, and separation of charge (S$_\text{c}$) confinement in the confinement region.   Our conjecture is that the custodial symmetry breaking transition, and the S$_\text{c}$-to-C confinement transition, coincide.  In the abelian model $S_{MGL}$ there
is also somewhere a transition between logarithmic and linear confinement.  Could it be that this transition is also
associated with spontaneous breaking of a custodial symmetry, which we now identify as a spin glass transition?

    At this point we must confront the Mermin-Wagner theorem, mentioned at the beginning of this section.  The modified
Ginzburg-Landau action $S_{MGL}$ differs from $S_{GL}$ in that the Higgs part of the action is a set of uncoupled actions in different $xy$-planes. As a result, the custodial symmetries $\phi(x) \ra e^{i\a(z)} \phi(x)$ can be regarded as a 
$U(1)\times U(1) \times ...\times U(1)$ symmetry, where each $U(1)$ factor is an independent global symmetry acting in a particular $xy$ plane at fixed $z$ and which could, in principle, break spontaneously. The $\Phi$ order parameter could detect such symmetry breaking in $Z_{spin}(U)$, if it exists.  But a true symmetry breaking transition of this kind would imply the breaking of a continuous $U(1)$ symmetry in two dimensions, and this is ruled out by the Mermin-Wagner theorem.

   Despite this fact, we may note that in real materials that are thought to be spin glasses, it is
even now not known whether the spin glass transition is a true thermodynamic transition, 
and whether the global symmetries are truly broken spontaneously in the spin glass phase \cite{Stein2011SpinGO}.  Spin glasses are, however, characterized by metastable states with extremely long relaxation times, and this is a property which we can investigate
numerically, even granting the fact that a true custodial symmetry breaking transition cannot exist in the system described by
$S_{MGL}$.  Keeping the background gauge field fixed, metastability can be observed in the time variable of a molecular dynamics simulation, or the time variable of Langevin evolution, or, as the most convenient choice, in terms of the number of lattice sweeps $n_s$, in a Monte Carlo simulation in which the scalar field is updated but the gauge field $U$ is fixed.  In this
last case metastability implies very long autocorrelation times.

    Our procedure, after initial thermalization, is to compute $\Phi$ via a ``Monte-Carlo-within-a-Monte Carlo'' simulation.  This means that we update the scalar and gauge fields together, according to the usual
Metropolis algorithm, for some number of sweeps (we chose one hundred). This generates a starting configuration, with the gauge field $U$ chosen
from the probability distribution \rf{PU}.  The data-taking sweep actually consists of $n_s$ sweeps, in which the scalar field
$\phi(x)$ is updated, but the gauge field $U_i(x)$ is held fixed.  At each data-taking sweep we compute the
average
\bea
        | \overline{\phi}|  = {1\over V} \sum_x   {1\over n_s} \left| \sum_{n=1}^{n_s} \phi(x,n) \right|
\eea
where index $n$ denotes the $n$-th Monte Carlo sweep with $U$ fixed.  Then averaging $|\overline{\phi}|$ over all
data-taking sweeps in the simulation gives us an estimate for $\Phi(n_s)$.  If there were a true transition, then on general statistical grounds we would expect
\beq
           \Phi(n_s) = \Phi_\infty + {c \over \sqrt{n_s}}
\label{Phi_n}
\eeq
with $c$ a constant, and $\Phi_\infty$ non-zero in the case of a true transition.  Because of the Mermin-Wagner theorem it must
be that $\Phi(n_s) \ra 0$ as $n_s \ra \infty$, even in the infinite volume limit.  But if there exist metastable states
with very long relaxation times, it could be that, beyond some line of critical couplings, the large volume data fits \rf{Phi_n} with $\Phi_\infty >0$ up to very large values of $n_s$.  This would indicate the existence of a line of quasi-transitions into a 
spin glass state of some kind,  which is a precursor to genuine transitions in higher spatial dimensions.

\begin{figure}[t!]
\centerline{\includegraphics[scale=0.8]{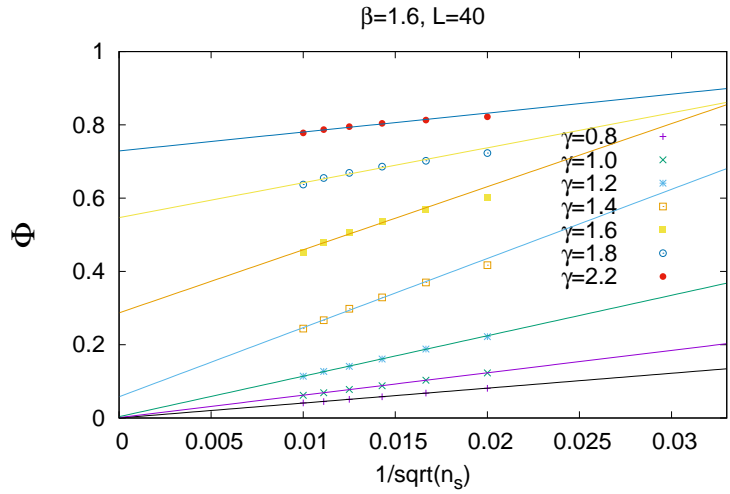}}
\caption{The custodial symmetry/spin glass order parameter $\Phi$ vs.\ $1/\sqrt{n_s}$, where
$n_s$ is the number of update sweeps of the scalar field with the gauge field held
fixed.  The data shown is at $\b=1.6$ and $\g$ values from 0.8 to 2.2 on a $40^3$ lattice volume. The straight lines
are best fits to the data at fixed $\b,\g$. There is an (apparent) transition at a critical coupling
$\g_c$ between $\g=1.2$ and 1.4, beyond which the extrapolated value of $\Phi$ at $n_s=\infty$ appears to be non-zero. } 
\label{b16}
\end{figure}

\begin{figure}[h!]
\centerline{\includegraphics[scale=0.8]{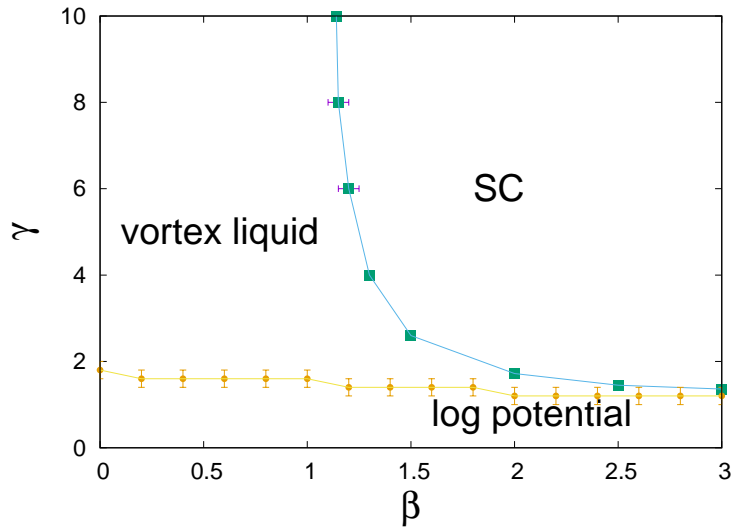}}
\caption{Location of the spin glass ``quasi-transition'' (yellow circles) in the phase diagram of the
modified Ginzburg-Landau lattice action.} 
\label{bg_glass}
\end{figure}

    In Fig.\ \ref{b16} we display our results for $\Phi(n_s)$ vs.\ $1/\sqrt{n_s}$, at $\b=1.6$, and various $\g$, computed
on a $40^3$ lattice volume.  Data points in the figure were computed at $n_s$ up to $n_s=10000$.  For $\g$ below the quasi-transition point at $1.2<\g_c<1.4$, the data falls on a straight line which extrapolates to 
$\Phi_\infty = 0$ at $n_s \ra \infty$.  Above the transition, the data appears to extrapolate to a non-zero value of $\Phi_\infty$.  By estimating at which $\g$ value, for fixed
$\b$, the extrapolated value for $\Phi_\infty$ begins to move away from zero, we arrive at the quasi-transition points
shown in Fig.\ \ref{bg_glass}.  

\begin{figure}[h!]
\centerline{\includegraphics[scale=0.8]{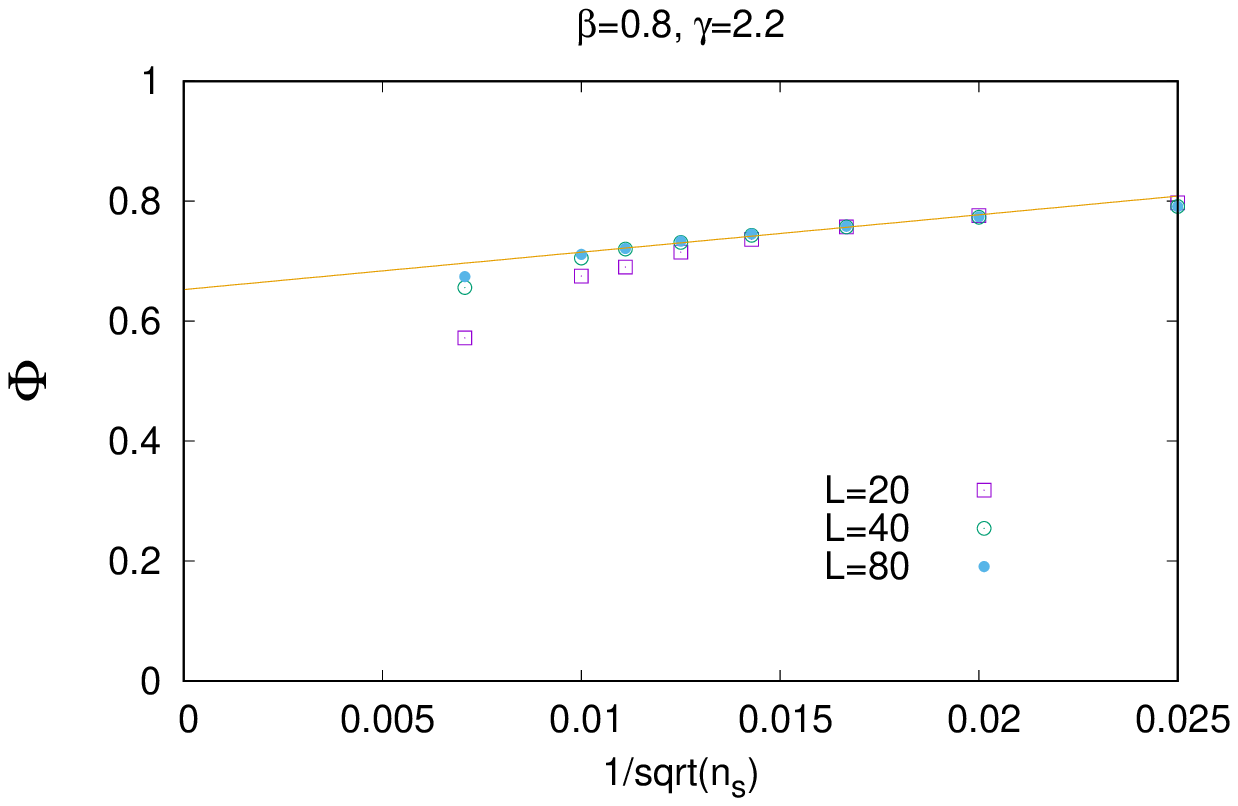}}
\caption{The spin glass order parameter $\Phi(n_s)$ vs.\ $1/\sqrt{n_s}$ at $\b=0.8, \g=2.2$, computed on lattice volumes $L^3$ with $L=20,40,80$.  In this case the simulations were carried out to $n_s=20,000$.} 
\label{nsym1}
\end{figure} 

    Of course, at $n_s \ra \infty$ we must have $\Phi_\infty \ra 0$, in a finite volume, even if there were a true transition in the infinite volume limit.  In Fig.\ \ref{nsym1} we show the data at $\b=0.8, \g=2.2$, which is inside the linearly confining
region which have argued corresponds to the pseudogap region in cuprates, on $20^3, 40^3, 80^3$ lattice volumes,
up to $n_s=20,000$ lattice sweeps.  At all three volumes we see the data fall away from the straight line (a fit to the
first few data points at low $n_s$),  presumably heading to zero at $n_s=\infty$.  The falloff is, however, most pronounced
at the smallest $20^3$ volumes, and the data points seem to tend upwards towards the straight line as the volume
increases.  Were it not for the Mermin-Wagner theorem, we would probably conclude that $\Phi_\infty$ is non-zero in the
infinite volume limit.  This cannot be true.  Nevertheless, there is no strong indication from the data that $\Phi_\infty$
vanishes at infinite volumes, and this implies the existence of states with very long relaxation times as measured by the $n_{spin}$, parameter, characteristic of some sort of spin glass phase, with a fairly abrupt transition, as $\g$ varies at fixed $\b$,  to a phase of this kind. We regard this as a precursor to the true custodial symmetry breaking transition expected in higher dimensions.  Our conjecture is that this quasi-transition is associated with a transition from the massless phase, i.e.\ logarithmic confinement, to the ``gapped'' phases, namely the pseudogap and superconducting phases.

\section{\label{sec6} Towards a realistic effective action}       
       
       In this article we have advocated the use of electromagnetic observables, i.e.\ Wilson loops, Polyakov loops, and $Z_2$ projected Wilson loops, as a probe of cuprate phase structure, and in particular we have argued that there are strong analogies between the pseudogap phase of the cuprates, and the confined phase of a non-abelian gauge theory.  However, we have illustrated the use of those observables in a theory which, while incorporating certain important features, is surely not a very realistic model of the cuprates.  For one thing, the momentum-space anisotropies associated with D-wave superconductivity are absent in this model, and so it goes wrong already at this early stage.   In addition, if we take e.g.\ 
${T=100~\mbox{K}}$ and $\b$  to be of O(1) we get a lattice spacing on the order of $10^{-4}$ m (see footnote 1), and this is far larger than the radius of a magnetic vortex in a cuprate, which is the scale where we might expect the effective theory to apply.  In this section we would like to indicate how one might derive, and solve numerically, a more realistic model.

      The obvious starting point is the Hubbard model Hamiltonian
\bea
      H &=& - {\bf q} \sum_x \sum_{i=1}^2 \Bigl( c^\dg_\s(x) c_\s(x+\ih ) + c^\dg_\s(x) c_\s(x-\ih) \Bigl)  \non \\
          & & + {\bf Y} \sum_x (n_\ua(x) - \oh)(n_\da(x) - \oh) - \overline{\bf \m} \sum_x (n_\ua(x)  + n_\da(x)) \non \\
          &=& H_K + H_V          
\label{Hub}
\eea       
where $H_K$ is the hopping term proportional to ${\bf q}$, $H_V$ contains the local terms proportional to
 ${\bf Y},\overline{\bf \m}$, and constants  ${\bf q},{\bf Y},\overline{\bf \m}$ all have units of energy, and
\beq
           Z(\b) = \tr \ e^{-\b H}    
\eeq
We have made the slight generalization that the sum over $x$ in \rf{Hub} is taken to be a sum over sites in a {\it three}-dimensional lattice.  Electrons hop only between nearest sites in planes parallel to the $x-y$ plane, so at this inital stage the introduction of numerous planes perpendicular to the $z$-axis is just a redundancy. 
 
   Lattice Monte Carlo treatments of the Hubbard model are generally based on the seminal work of Blankenbecler, Scalapino, and Sugar (BSS)  \cite{Blankenbecler:1981jt} (see e.g.\ the early studies in \cite{Hirsch,White:1989zz}, and more recent investigations in \cite{Smith:2014tha,Beyl:2017kwp,Buividovich:2018yar,huang2018strange,Ulybyshev:2019hfm,Fukuma:2019wbv}), and we will also follow along these lines.  In order to address the questions we are interested in we have to modify the model to include the electromagnetic field, and also explain how we would deal with the sign problem, which is inevitable away from half-filling.  The electromagnetic field is introduced in a fairly obvious way, by following
along the BSS derivation and then imposing local gauge invariance.  We suggest here that the sign problem
might be addressed by the complex Langevin approach.  We should stress that this section does not contain any numerical results; it is only a proposal.  The calculation that we outline here is computationally intensive, and is reserved 
for future work.

    Starting from
\bea
    Z =  \tr \ e^{-\b H} = \tr \left(e^{-\d t H} \right)^{N_t} \approx \tr \left(e^{-\d t H_K} e^{-\d t H_V} \right)^{N_t}
\label{partfun}
\eea
where $\d t = \b/N_t$, we define the dimensionless constants
\beq
          q = \b {\bf q} ~~,~~ Y = \b {\bf Y} ~~,~~ \m = \b \overline{\bf \m}
\eeq
Then re-express the term proportional to $Y$
via the Hubbard-Stratonovich transformation
\bea
     & &   \exp\left[-{1\over N_t}Y(n_\ua(x) - \oh)(n_\da(x) - \oh) \right] = \left({1\over N_t \pi}\right)^\oh e^{-Y/4N_t} \non \\
     & & \qquad \qquad  \times  \int d \phi(x) \exp\left[ -{1\over N_t} ( \phi^2(x) + \sqrt{2Y}(n_\ua(x)  - n_\da(x))\right]
\eea
It was shown by BSS that the partition function \rf{partfun} can be rewritten as a Euclidean time path integral over bosonic
($\phi$) and Grassman ($\pbar,\psi$) valued fields
\beq
    Z = \int \prod_{t=1}^{N_t} \prod_x d\pbar_\s(x,t) d\psi_\s(x,t) d\phi(x,t)  ~ e^{-S}
\eeq
where
\bea
    S &=& {1\over N_t} \sum_{x,t} \phi^2(x,t) +      
            \sum_{x,x'}\sum_{t,t'} \sum_{\s=\ua,\da} \pbar_\s(x,t) M_\s(xt,x't') \psi_\s(x',t')
\label{BSS}
\eea
with
\bea
M_\s = \left[ \begin{array}{ccccc}
             \mathbbm{1} & 0 & 0 & \cdot  \cdot  \cdot & B^\s_1 \cr
                 -B^\s_2  & \mathbbm{1}  & 0 & \cdot  \cdot  \cdot & 0 \cr
                 0 & -B^\s_3  & \mathbbm{1} & \cdot  \cdot  \cdot & 0 \cr
                 \cdot & \cdot & \cdot & & 0 \cr
                 \cdot & \cdot & \cdot & & 0 \cr
                 \cdot & \cdot & \cdot & & 0 \cr
                 \cdot & \cdot &\cdot  \cdot  \cdot & -B^\s_{N_t} & \mathbbm{1}  \end{array} \right]
\eea
In this expression the time index $t'$ runs left to right, index $t$ runs along the vertical,
$\mathbbm{1}$ and $B^\s_t$ are matrices of dimension $N_S \times N_S$, where $N_S$ is the total
number of lattice sites, and where
\beq
          B_t^\s = e^{-K/N_t} e^{-V_s/N_t}
\eeq
with
\bea
          K(x,y) &=& -q \sum_{i=1,2} (\d_{x+\ih,y} + \d_{x-\ih,y}) \non \\
          V_\s(x,y) &=& \d_{xy} ( \m + s_\s \sqrt{2Y} \phi(x,t) ) ~~\text{where}~~
          s_\s = \{ \begin{array}{cl}
                          +1 & \s = \ua \cr
                          -1 &  \s = \da \end{array}
\eea
Integrating over the Grassman variables leads to an effective action for the Hubbard-Stratonovich field
\bea
            Z &=&  \int \prod_{t=1}^{N_t} \prod_x d\phi(x,t) \det[M_\ua] \det[M_\da] 
              \exp\left[ -{1\over N_t} \sum_{t=1}^{N_t} \sum_x \phi^2(x,t) \right] \non \\
              &=& \int \prod_{t=1}^{N_t} \prod_x d\phi(x,t) e^{-S_\phi}
\eea
where
\bea
        S_\phi = {1\over N_t} \sum_{t=1}^{N_t} \sum_x \phi^2(x,t) - \tr \ln M_\ua - \tr \ln M_\da
\label{Sf}
\eea
In the case of half filling it can be shown that $\det[M_\ua]$ and $\det[M_\da]$ have the same sign  \cite{Hirsch},
so there is no sign problem in that situation.  So far this is all standard.

    The next step is to promote the global invariance of \rf{BSS} under $\psi \ra e^{i\a} \psi$, ${\pbar \ra e^{-i\a} \pbar}$
to a local invariance, keeping in mind that the $N_S \times N_S$ matrix $B_t$ connects $\pbar$ at time $t$ with
$\psi$ at time $t-1$.  Local invariance is achieved by first modifying the hopping term
\beq 
K(x,y) \ra K_U(x,y) = -q \sum_{i=1,2} (\d_{x+\ih,y}U_i(x,t) + \d_{x-\ih,y} U^\dg_i(x-\ih)) 
\eeq
and then, since $M_\s$ connects $\psi$ at time $t-1$ to $\pbar$ at time $t$, we introduce the diagonal matrix 
${\cal U}_t$ with matrix elements
\beq
          [{\cal U}_t]_{xy} = \d_{xy} U^\dg_4(x,t-1)
\eeq
and redefine
\beq
       B_t^\s = {\cal U}_t e^{-K_U/N_t} e^{-V_\s/N_t} 
\label{Bmod}
\eeq
We then add the latticized action $S_W$ for the electromagnetic field to the effective action.  Here we
must take account of the different lattice spacings that are involved.  The Hubbard model assumes that
electrons hop in a plane, between points separated by some spacing $a$.  The neighboring planes are separated
by a distance $a_z$, with $a$ and $a_z$ at the interatomic distance scale, presumed to be fixed (like ${\bf q}, {\bf Y}$) by comparison to some real material.  The Euclidean time step is $\d t = \b/N_t$, with continuous time obtained in the
$N_t \ra \infty$ limit.  With gauge link variables
$U_\m(x) = \exp[i\th_\m(x)]$ and $\th_{\m \n}$ defined in \rf{thmunu}, the Wilson lattice action for this asymmetric lattice, having the appropriate continuum limit, is
\bea
S_W &=& -{1\over e^2} \sum_x \bigg\{ {N_t \over \b} \left( a_z(\cos \th_{14} + \cos \th_{24})
                   + {a^2 \over a_z} \cos \th_{34} \right) \non \\
          & & + {\b \over N_t}\left( {1\over a_z} (\cos \th_{13} + \cos\th_{23}) +
                    {a_z \over a^2} \cos \th_{12} \right)\bigg\}
\eea
The final effective action is 
\beq
      S_{eff}(\th_\m,\phi) = S_\phi(\th_\m,\phi) + S_W(\th_\m)
\label{Seff}
\eeq 
where $S_\phi$ is the action in \rf{Sf} with the modification \rf{Bmod}.  For arbitrary chemical potential there is certainly
a sign problem.

   We are not aware of any numerical treatment of the Hubbard model which incorporates
a coupling to the quantized electromagnetic field, probably because this field is considered extraneous
to the underlying physics.  The experimental evidence of vortex effects in the pseudogap phase of cuprates suggests that this coupling should not be ignored. 
However, with or without the coupling to the electromagnetic field, one must
confront the sign problem in regions of interest in the cuprate phase diagram.  There are a number of
approaches in the literature.  One idea, which goes back to \cite{White:1989zz}, is to simulate the theory using the
modulus $|\det M_\ua \det M_\da|$.  This is known as the ``sign-quenched'' technique in the particle 
physics literature, and it is known to go wrong for QCD at high baryon density \cite{Ipsen:2012ug}.  Three other methods which have been studied extensively are the complex Langevin equation \cite{Aarts:2013uxa,Sexty:2013ica,berger2019complex}, the thimble approach
\cite{Cristoforetti:2013wha}, and the LLR algorithm \cite{Langfeld:2016kty}.  None of these methods is perfect.  Each has been shown to work successfully in some models, and to fail in others.   The thimble approach has recently been applied to the two-dimensional Hubbard model \cite{Ulybyshev:2019hfm,Fukuma:2019wbv}, but thus far only for tiny ($2\times 2$) lattices.

    We propose to apply the complex Langevin equation to the effective Hubbard model action which includes the
quantized gauge field.  Discretizing the fictitious Langevin time $\t$, the Langevin equations are
\bea
          \phi(x,t,\t+\e) &=&  -\left[{\pa S_{eff} \over \pa \phi(x,t)}\right]_\t \e + \eta_\phi(x,t,\t) \sqrt{\e} \non \\
          \th_\m(x,t,\t+\e) &=&  -\left[{\pa S_{eff} \over \pa \th_\m(x,t)}\right]_\t \e + \eta_{\th_\m}(x,t,\t) \sqrt{\e} 
\label{Lang}
\eea
where the notation $[...]_\t$ means that the quantity in brackets is to be evaluated with fields at Langevin time $\t$,
and the $\eta$ fields are random numbers with a probability distribution such that  
\bea
     \langle \eta_\phi(x,t,\t) \eta_\phi(x',t',\t') \rangle &=& 2 \d_{xx'} \d_{tt'} \d_{\t\t'} \non \\
     \langle \eta_{\th_\m}(x,t,\t) \eta_{\th_\n}(x',t',\t') \rangle &=& 2 \d_{\m\n} \d_{xx'} \d_{tt'} \d_{\t\t'} \non \\    
     \langle \eta_\phi(x,t,\t) \eta_{\th_\m}(x',t',\t') \rangle &=& 0
\eea
Since the effective action $S_{eff}$ is in general complex away from half-filling, these equations are only consistent if the $\phi, \th_\m$ fields are also complex (and of course this means that gauge link variables are no longer unimodular), 
but the $\eta$ fields are always taken to be real.   Solving these equations numerically, and averaging observables over the fictitious Langevin time, is the essence of the complex Langevin approach.  

    The trace of logarithms in the effective action give rise to
\bea
   - {\pa S_{eff} \over \pa \phi(x,t)} &=& - {2\over N_t} \phi(x,t) 
               + \sum_{\s=\ua,\da} \tr \left[M^{-1}_\s {\pa \over \pa \phi(x,t)} M_\s \right]  \non \\
   - {\pa S_{eff} \over \pa \th_\m(x,t)} &=& - {\pa S_W \over \pa \th_\m(x,t)} 
               + \sum_{\s=\ua,\da}  \tr \left[M^{-1}_\s {\pa \over \pa \th_\m(x,t)} M_\s \right]
\eea
Since $M_\s$ has dimensions of $N\times N$, where $N=N_t N_S$ is the number of lattice sites, direct inversion
of $M_\s$ at each Langevin time step is impractical
if $N$ is large, but there are tricks which avoid direct inversion, as explained in, e.g., 
\cite{Sexty:2013ica}.  The idea is to approximate the trace with a stochastic $N$-component noise vector ${\bf v}$
\beq
        \tr \left[ M^{-1}_\s \pa M_\s \right] \approx {\bf v}^\dg \cdot M^{-1}_\s \pa M_\s {\bf v} 
\eeq
where ${\bf v}$ is drawn from a gaussian probability distribution such that
\beq
          \langle v_i^* v_j \rangle = \d_{ij}
\eeq
One must then solve the linear system 
\beq 
M^\dg_\s{\bf w} = {\bf v}
\label{linear}
\eeq 
after which the trace becomes
\beq
       \tr \left[M^{-1}_\s \pa M_\s \right] \approx {\bf w}^\dg  \cdot \pa M_\s {\bf v}
\eeq

   The most computationally intensive part of this algorithm is the solution of the linear system of equations \rf{linear}.
This is practical if the matrix $M_\s$ is sparse, which is not quite true of the $N_S \times N_S$ matrix $B^\s_t$.  The problem
is that while the matrix $K_U$ is sparse, the exponential of this matrix is not.  One possibility is to expand $\exp[-K_U/N_t]$
to first or second order in $1/N_t$, in which case $M_\s$ will be a sparse matrix.  There are indications that
expansion to first order may not be sufficient, cf.\ \cite{Buividovich:2018yar} unless $N_t$ is quite large,, and this reference also presents a method (the ``Schur complement solver''), which speeds up the solution of the linear system without approximating the exponential $\exp[-K_U/N_t]$.  The method has also been applied in the thimble approach of \cite{Ulybyshev:2019hfm}.  Alternatively, one might simply expand $\exp[-K_U/N_t]$ to second order in $1/N_t$.

    By these means it ought to be possible implement evolution in Langevin time and to compute gauge field observables, in particular Wilson loops, Polyakov loops, vortex densities, electromagnetic field strengths, and so on. The complex Langevin method is practical, and supplies
numerical answers to numerical questions.  What is not certain is whether those answers are correct.  The validity of
the method is not guaranteed when the action is non-holomorphic \cite{Mollgaard:2013qra,Aarts:2017vrv}, which is the situation for the Hubbard model.  Sometimes
the complex Langevin approach will supply the correct answers for non-holomorphic actions, and sometimes not
\cite{Greensite:2014cxa}.  There are certain tests for validity; one can only try and see.

     The effective action $S_{eff}$ in \rf{Seff}, unlike our modified Ginzburg-Landau model \rf{MGL}, is not a gauge-Higgs theory in the usual sense because the scalar Hubbard-Stratonovich field is uncharged.  If we are only interested in
gauge field observables, this is not a problem.  It has been suggested (in the context of graphene) that effective theories
involving the charge-neutral Hubbard-Stratonovich field (but not the gauge field) can be used to detect charge and spin density wave order \cite{Buividovich:2018yar}.  On the other hand, a theory with a neutral scalar field cannot be used
to investigate the possible spin-glass nature of the Higgs and pseudogap phases, as suggested in the last section.  So the
question is whether one could follow the steps in the derivation of the Ginzburg-Landau theory from the BCS Hamiltonian, as presented in many textbooks (e.g.\ \cite{altland_simons_2010}), using a charged rather than a neutral Hubble-Stratonovich field to decouple the four-fermi interaction term.  The problem, however, is that unlike the BCS Hamiltonian, where the four-fermi attraction is attractive, in the Hubbard model the four-fermi term is repulsive, and the opposite sign of this term as compared to the BCS theory results in a ``wrong-sign'' in front of the term $\phi^*(x) \phi(x)$ quadratic in the Hubbard-Stratonovich field, i.e.\ if one could argue that the integral
\beq
          \int d\phi \exp[ (\phi^* - \sqrt{Y} c^\dg_\ua c^\dg_\da)(\phi - \sqrt{Y} c_\da c_\ua)/N_t]
\eeq
was meaningful, then the Hubbard-Stratonovich trick would replace the four-fermi term in the Hubbard model Hamiltonian by
the identity
\bea
 & &  - (\phi^* - \sqrt{Y} c^\dg_\ua c^\dg_\da)(\phi - \sqrt{Y} c_\da c_\ua) + Y c^\dg_\ua(x) c_\ua(x) c^\dg_\da(x) c_\da(x) \non \\
  & &  \qquad =
    - \phi^*(x) \phi(x) + \sqrt{Y} \phi^*(x) c_\da(x) c_\ua(x) + \phi(x) \sqrt{Y} c^\dg_\ua(x) c^\dg_\da(x)
\eea
with the obvious drawback that the integration over $\phi$ in $e^{-\b H}$ is exponentially divergent.
The situation is not necessarily hopeless.  Wrong-sign quadratic terms have been encountered elsewhere, notably in
Euclidean quantum gravity, where the kinetic term of the metric conformal factor has the wrong sign, and the recommended prescription \cite{Gibbons:1978ac} is to interpret expectation values of this field via a deformation of the contour of
integration in the path integral.  According to this prescription one would have, e.g.
\beq
           \langle x^2 \rangle = {\int dx \ x^2 e^{+x^2} \over  \int dx \ e^{+x^2} } = -\oh
\eeq
Another approach to so-called
``bottomless action'' theories, where the value of the Euclidean action is unbounded from below, is presented in 
\cite{Greensite:1983yc}.  But of course the difficulties of the wrong-sign term, coupled with the already thorny technicalities
of the sign problem associated with the chemical potential, suggests that attention should be directed first to the
effective theory with a neutral Hubbard-Stratonovich term, and a conventional sign in front of the quadratic term in
the effective action.

     Perhaps the greatest advantage of focussing on electromagnetic observables, i.e.\ Wilson loops, Polyakov lines, and center-projected loops is that they allow us to study numerically the electromagnetic properties of the theory in the superconducting, pseudogap, and other phases of the theory even if the effective theory involves only a neutral Hubbard-Stratonovich field, which may be the only practical possibility.

\section{\label{sec7} Conclusions}

   We have presented a modified lattice version of the time-independent Ginzburg-Landau model, containing a compact U(1) gauge field with a no-monopole constraint, and an action for the scalar field with nearest-neighbor couplings limited to two dimensional $x-y$ planes of the three dimensional volume.   In this model we detect in numerical simulations
\begin{enumerate}
\item an area-law falloff for Wilson loops in
the x-y planes, in regions of small $\b$ and large $\g$;
\item a superconducting phase at large $\b$ and large $\g$;
\item a falloff consistent with a logarithmic potential at large $\b$ and
small $\g$.
\end{enumerate}
The superconducting phase is distinguished from other regions of the phase diagram by the spontaneous breaking
of a global Z$_\text{2}$  symmetry.

In the region of area-law falloff we have used a Z$_\text{2}$ 
projection procedure to identify the location of vortex configurations, and provided evidence that the identified vortices are
responsible for the area-law falloff.  This ties in to the center vortex theory of confinement in non-abelian gauge theories,  as briefly reviewed here.  It is interesting that this is an example where the introduction of a matter field can induce an area-law falloff in a U(1) gauge theory with a no-monopole constraint, in which the area-law
falloff would otherwise be absent.

    While we do not suggest that the modified lattice Ginzburg-Landau model studied here is a realistic model of the physics of cuprates, we do believe that it furnishes an example of a superconducting to vortex liquid phase in the context of a U(1) gauge theory, in which the modulus of the scalar field is a non-zero constant. This may be relevant to the superconducting to pseudogap transition found in the cuprates.

   On the experimental side, we note that there have, in fact, been measurements of magnetic susceptibility along a planar area of a thin film of cuprate material in the superconducting phase, e.g.\  \cite{Moler}.  Perhaps it is also feasible to compute the expectation value of Wilson loops of fixed area in the pseudogap phase. This could be accomplished by measuring the magnetic flux $\Phi(t)$ through a fixed loop $C$ as a function of time, from which we derive the 
corresponding Wilson loop observable $\exp[ie\Phi(t)/\hbar]$, also as a function of time, and then averaging with respect to time to determine $W(C)$.  The behavior of Wilson loops is fundamental to our understanding of the strong nuclear force, so their experimental determination in cuprate materials is an intriguing possibility.

   We finally wish to stress the utility of gauge field observables in the numerical simulation of more realistic theoretical models of the cuprates, such as the Hubbard model.  This calls for introducing the electromagnetic gauge field in such models, as discussed in the previous section.  There is no need, in this approach, to have a charged order parameter associated with pairing, e.g.\ a charged Hubbard-Stratanovich field in the effective action.   The distinction between the normal, superconducting, and pseudogap phases can be made entirely in terms of gauge field observables, i.e.\ the Wilson, Polyakov, and $Z_2$-projected loops, and the challenge is mainly to confront the sign problem, perhaps via the 
complex Langevin equation, or by some other means.

\bigskip

\ni {\bf Acknowledgements} 

 We thank Makoto Hashimoto for discussions.  JG's research is supported by the U.S.\ Department of Energy under Grant No.\ DE-SC0013682.  


\bibliography{hitc}

\begin{thebibliography}{10}
\expandafter\ifx\csname url\endcsname\relax
  \def\url#1{\texttt{#1}}\fi
\expandafter\ifx\csname urlprefix\endcsname\relax\def\urlprefix{URL }\fi
\expandafter\ifx\csname href\endcsname\relax
  \def\href#1#2{#2} \def\path#1{#1}\fi

\bibitem{2006PhRvB..73b4510W}
Y.~{Wang}, L.~{Li}, N.~P. {Ong}, {Nernst effect in high- T$_{c}$
  superconductors}, Phys. Rev. B 73~(2) (2006) 024510.
\newblock \href {http://arxiv.org/abs/cond-mat/0510470}
  {\path{arXiv:cond-mat/0510470}}.

\bibitem{2007JMMM..310..460L}
L.~{Li}, Y.~{Wang}, M.~J. {Naughton}, S.~{Komiya}, S.~{Ono}, Y.~{Ando}, N.~P.
  {Ong}, {Magnetization, Nernst effect and vorticity in the cuprates}, Journal
  of Magnetism and Magnetic Materials 310 (2007) 460--466.
\newblock \href {http://arxiv.org/abs/cond-mat/0611731}
  {\path{arXiv:cond-mat/0611731}}.

\bibitem{2010PhRvB..81e4510L}
L.~{Li}, Y.~{Wang}, S.~{Komiya}, S.~{Ono}, Y.~{Ando}, G.~D. {Gu}, N.~P. {Ong},
  {Diamagnetism and Cooper pairing above T$_{c}$ in cuprates}, Phys. Rev. B
  81~(5) (2010) 054510.
\newblock \href {http://arxiv.org/abs/0906.1823} {\path{arXiv:0906.1823}}.

\bibitem{2018arXiv180411186A}
P.~W. {Anderson}, {Four Last Conjectures}, ArXiv e-prints\href
  {http://arxiv.org/abs/1804.11186} {\path{arXiv:1804.11186}}.

\bibitem{2016arXiv161203919A}
P.~W. {Anderson}, {Last Words on the Cuprates}, ArXiv e-prints\href
  {http://arxiv.org/abs/1612.03919} {\path{arXiv:1612.03919}}.

\bibitem{Emery}
V.~J. Emery, S.~A. Kivelson, Superconductivity in bad metals, Phys. Rev. Lett.
  74 (1995) 3253--3256.

\bibitem{Sachdev}
S.~Sachdev, H.~D. Scammell, M.~S. Scheurer, G.~Tarnopolsky,
  \href{https://link.aps.org/doi/10.1103/PhysRevB.99.054516}{Gauge theory for
  the cuprates near optimal doping}, Phys. Rev. B 99 (2019) 054516.
\newblock \href {http://dx.doi.org/10.1103/PhysRevB.99.054516}
  {\path{doi:10.1103/PhysRevB.99.054516}}.
\newline\urlprefix\url{https://link.aps.org/doi/10.1103/PhysRevB.99.054516}

\bibitem{DeGrand:1980eq}
T.~A. DeGrand, D.~Toussaint, {Topological Excitations and Monte Carlo
  Simulation of Abelian Gauge Theory}, Phys. Rev. D22 (1980) 2478,
  [,194(1980)].

\bibitem{Mack:1979gb}
G.~Mack, V.~B. Petkova, {Z2 Monopoles in the Standard SU(2) Lattice Gauge
  Theory Model}, Z. Phys. C12 (1982) 177.

\bibitem{Elitzur:1975im}
S.~Elitzur, {Impossibility of Spontaneously Breaking Local Symmetries},
  Phys.Rev. D12 (1975) 3978--3982.

\bibitem{Osterwalder:1977pc}
K.~Osterwalder, E.~Seiler, {Gauge Field Theories on the Lattice}, Annals Phys.
  110 (1978) 440.

\bibitem{Fradkin:1978dv}
E.~H. Fradkin, S.~H. Shenker, {Phase Diagrams of Lattice Gauge Theories with
  Higgs Fields}, Phys.Rev. D19 (1979) 3682--3697.

\bibitem{Caudy:2007sf}
W.~Caudy, J.~Greensite, {On the ambiguity of spontaneously broken gauge
  symmetry}, Phys.Rev. D78 (2008) 025018.
\newblock \href {http://arxiv.org/abs/0712.0999} {\path{arXiv:0712.0999}}.

\bibitem{Willenbrock:2004hu}
S.~Willenbrock, {Symmetries of the standard model, }\href
  {http://arxiv.org/abs/hep-ph/0410370} {\path{arXiv:hep-ph/0410370}}.

\bibitem{Maas:2019nso}
A.~Maas, {Brout-Englert-Higgs physics: From foundations to phenomenology},
  Prog. Part. Nucl. Phys. 106 (2019) 132--209.
\newblock \href {http://dx.doi.org/10.1016/j.ppnp.2019.02.003}
  {\path{doi:10.1016/j.ppnp.2019.02.003}}.

\bibitem{Greensite:2018mhh}
J.~Greensite, K.~Matsuyama, {What symmetry is actually broken in the Higgs
  phase of a gauge-Higgs theory?}, Phys. Rev. D98~(7) (2018) 074504.
\newblock \href {http://arxiv.org/abs/1805.00985} {\path{arXiv:1805.00985}}.

\bibitem{tHooft:1977nqb}
G.~'t~Hooft, {On the Phase Transition Towards Permanent Quark Confinement},
  Nucl. Phys. B138 (1978) 1--25.

\bibitem{Greensite:2011zz}
J.~Greensite, {An introduction to the confinement problem}, Springer Lect.
  Notes Phys. 821 (2011) 1--211.

\bibitem{Greensite:2003bk}
J.~Greensite, {The Confinement problem in lattice gauge theory}, Prog. Part.
  Nucl. Phys. 51 (2003) 1.
\newblock \href {http://arxiv.org/abs/hep-lat/0301023}
  {\path{arXiv:hep-lat/0301023}}.

\bibitem{Engelhardt:1998wu}
M.~Engelhardt, K.~Langfeld, H.~Reinhardt, O.~Tennert, {Interaction of confining
  vortices in SU(2) lattice gauge theory}, Phys. Lett. B431 (1998) 141--146.
\newblock \href {http://arxiv.org/abs/hep-lat/9801030}
  {\path{arXiv:hep-lat/9801030}}.

\bibitem{Kamleh:2017thj}
W.~Kamleh, D.~Leinweber, D.~Trewartha, {Centre Vortices As The Origin Of Quark
  Confinement}, PoS INPC2016 (2017) 293.

\bibitem{Trewartha:2017ive}
D.~Trewartha, W.~Kamleh, D.~Leinweber, {Centre vortex removal restores chiral
  symmetry}\href {http://arxiv.org/abs/1708.06789} {\path{arXiv:1708.06789}}.

\bibitem{Greensite:2017ajx}
J.~Greensite, K.~Matsuyama, {Confinement criterion for gauge theories with
  matter fields}, Phys. Rev. D96~(9) (2017) 094510.
\newblock \href {http://arxiv.org/abs/1708.08979} {\path{arXiv:1708.08979}},
  \href {http://dx.doi.org/10.1103/PhysRevD.96.094510}
  {\path{doi:10.1103/PhysRevD.96.094510}}.

\bibitem{Polyakov:1976fu}
A.~M. Polyakov, {Quark Confinement and Topology of Gauge Groups}, Nucl. Phys.
  B120 (1977) 429--458.

\bibitem{duncan2012}
A.~Duncan, \href{https://books.google.com/books?id=MuH0TQvpY5sC}{The Conceptual
  Framework of Quantum Field Theory}, EBSCO ebook academic collection, OUP
  Oxford, 2012.
\newline\urlprefix\url{https://books.google.com/books?id=MuH0TQvpY5sC}

\bibitem{GREITER2005217}
M.~Greiter, Is electromagnetic gauge invariance spontaneously violated in
  superconductors?, Annals of Physics 319~(1) (2005) 217 -- 249.
\newblock \href {http://dx.doi.org/https://doi.org/10.1016/j.aop.2005.03.008}
  {\path{doi:https://doi.org/10.1016/j.aop.2005.03.008}}.

\bibitem{schakel1998boulevard}
A.~M. Schakel, Boulevard of broken symmetries, arXiv preprint cond-mat/9805152.

\bibitem{dutch}
J.~{van Wezel}, J.~{van den Brink}, {Spontaneous symmetry breaking and
  decoherence in superconductors}, Phys. Rev. B77~(6) (2008) 064523.
\newblock \href {http://arxiv.org/abs/0706.1922} {\path{arXiv:0706.1922}},
  \href {http://dx.doi.org/10.1103/PhysRevB.77.064523}
  {\path{doi:10.1103/PhysRevB.77.064523}}.

\bibitem{Kennedy:1986ut}
T.~Kennedy, C.~King, {Spontaneous Symmetry Breakdown in the Abelian Higgs
  Model}, Commun. Math. Phys. 104 (1986) 327--347.
\newblock \href {http://dx.doi.org/10.1007/BF01211599}
  {\path{doi:10.1007/BF01211599}}.

\bibitem{Hansson:2004wca}
T.~H. Hansson, V.~Oganesyan, S.~L. Sondhi, {Superconductors are topologically
  ordered}, Annals Phys. 313~(2) (2004) 497--538.
\newblock \href {http://dx.doi.org/10.1016/j.aop.2004.05.006}
  {\path{doi:10.1016/j.aop.2004.05.006}}.

\bibitem{Edward_Anderson}
S.~F. Edwards, P.~W. Anderson, {Theory of spin glasses}, Journal of Physics F:
  Metal Physics 5~(5) (1975) 965--974.

\bibitem{Stein2011SpinGO}
D.~L. Stein, C.~M. Newman, Spin glasses: Old and new complexity, Complex
  Systems 20.

\bibitem{Blankenbecler:1981jt}
R.~Blankenbecler, D.~J. Scalapino, R.~L. Sugar, {Monte Carlo Calculations of
  Coupled Boson - Fermion Systems. 1.}, Phys. Rev. D24 (1981) 2278.

\bibitem{Hirsch}
J.~E. Hirsch, {Two-dimensional Hubbard model: Numerical simulation study},
  Phys. Rev. B31 (1985) 4403--4419.

\bibitem{White:1989zz}
S.~R. White, D.~J. Scalapino, R.~L. Sugar, E.~Y. Loh, J.~E. Gubernatis, R.~T.
  Scalettar, {Numerical study of the two-dimensional Hubbard model}, Phys. Rev.
  B40 (1989) 506--516.
\newblock \href {http://dx.doi.org/10.1103/PhysRevB.40.506}
  {\path{doi:10.1103/PhysRevB.40.506}}.

\bibitem{Smith:2014tha}
D.~Smith, L.~von Smekal, {Monte-Carlo simulation of the tight-binding model of
  graphene with partially screened Coulomb interactions}, Phys. Rev. B89~(19)
  (2014) 195429.
\newblock \href {http://arxiv.org/abs/1403.3620} {\path{arXiv:1403.3620}},
  \href {http://dx.doi.org/10.1103/PhysRevB.89.195429}
  {\path{doi:10.1103/PhysRevB.89.195429}}.

\bibitem{Beyl:2017kwp}
S.~Beyl, F.~Goth, F.~F. Assaad, {Revisiting the Hybrid Quantum Monte Carlo
  Method for Hubbard and Electron-Phonon Models}, Phys. Rev. B97~(8) (2018)
  085144.
\newblock \href {http://arxiv.org/abs/1708.03661} {\path{arXiv:1708.03661}},
  \href {http://dx.doi.org/10.1103/PhysRevB.97.085144}
  {\path{doi:10.1103/PhysRevB.97.085144}}.

\bibitem{Buividovich:2018yar}
P.~Buividovich, D.~Smith, M.~Ulybyshev, L.~von Smekal, {Hybrid-Monte-Carlo
  study of competing order in the extended fermionic Hubbard model on the
  hexagonal lattice}, Phys. Rev. B98~(23) (2018) 235129.
\newblock \href {http://arxiv.org/abs/1807.07025} {\path{arXiv:1807.07025}},
  \href {http://dx.doi.org/10.1103/PhysRevB.98.235129}
  {\path{doi:10.1103/PhysRevB.98.235129}}.

\bibitem{huang2018strange}
E.~W. Huang, R.~Sheppard, B.~Moritz, T.~P. Devereaux, Strange metallicity in
  the doped hubbard model (2018).
\newblock \href {http://arxiv.org/abs/1806.08346} {\path{arXiv:1806.08346}}.

\bibitem{Ulybyshev:2019hfm}
M.~Ulybyshev, C.~Winterowd, S.~Zafeiropoulos, {Taming the sign problem of the
  finite density Hubbard model via Lefschetz thimbles}\href
  {http://arxiv.org/abs/1906.02726} {\path{arXiv:1906.02726}}.

\bibitem{Fukuma:2019wbv}
M.~Fukuma, N.~Matsumoto, N.~Umeda, {Applying the tempered Lefschetz thimble
  method to the Hubbard model away from half-filling}\href
  {http://arxiv.org/abs/1906.04243} {\path{arXiv:1906.04243}}.

\bibitem{Ipsen:2012ug}
J.~R. Ipsen, K.~Splittorff, {Baryon Number Dirac Spectrum in QCD}, Phys. Rev.
  D86 (2012) 014508.
\newblock \href {http://arxiv.org/abs/1205.3093} {\path{arXiv:1205.3093}},
  \href {http://dx.doi.org/10.1103/PhysRevD.86.014508}
  {\path{doi:10.1103/PhysRevD.86.014508}}.

\bibitem{Aarts:2013uxa}
G.~Aarts, L.~Bongiovanni, E.~Seiler, D.~Sexty, I.-O. Stamatescu, {Controlling
  complex Langevin dynamics at finite density}, Eur. Phys. J. A49 (2013) 89.
\newblock \href {http://arxiv.org/abs/1303.6425} {\path{arXiv:1303.6425}},
  \href {http://dx.doi.org/10.1140/epja/i2013-13089-4}
  {\path{doi:10.1140/epja/i2013-13089-4}}.

\bibitem{Sexty:2013ica}
D.~Sexty, {Simulating full QCD at nonzero density using the complex Langevin
  equation}, Phys. Lett. B729 (2014) 108--111.
\newblock \href {http://arxiv.org/abs/1307.7748} {\path{arXiv:1307.7748}},
  \href {http://dx.doi.org/10.1016/j.physletb.2014.01.019}
  {\path{doi:10.1016/j.physletb.2014.01.019}}.

\bibitem{berger2019complex}
C.~E. Berger, L.~Rammelmüller, A.~C. Loheac, F.~Ehmann, J.~Braun, J.~E. Drut,
  Complex langevin and other approaches to the sign problem in quantum
  many-body physics (2019).
\newblock \href {http://arxiv.org/abs/1907.10183} {\path{arXiv:1907.10183}}.

\bibitem{Cristoforetti:2013wha}
M.~Cristoforetti, F.~Di~Renzo, A.~Mukherjee, L.~Scorzato, {Monte Carlo
  simulations on the Lefschetz thimble: Taming the sign problem}, Phys. Rev.
  D88~(5) (2013) 051501.
\newblock \href {http://arxiv.org/abs/1303.7204} {\path{arXiv:1303.7204}},
  \href {http://dx.doi.org/10.1103/PhysRevD.88.051501}
  {\path{doi:10.1103/PhysRevD.88.051501}}.

\bibitem{Langfeld:2016kty}
K.~Langfeld, {Density-of-states}, PoS LATTICE2016 (2017) 010.
\newblock \href {http://arxiv.org/abs/1610.09856} {\path{arXiv:1610.09856}},
  \href {http://dx.doi.org/10.22323/1.256.0010}
  {\path{doi:10.22323/1.256.0010}}.

\bibitem{Mollgaard:2013qra}
A.~Mollgaard, K.~Splittorff, {Complex Langevin Dynamics for chiral Random
  Matrix Theory}, Phys.Rev. D88 (2013) 116007.
\newblock \href {http://arxiv.org/abs/1309.4335} {\path{arXiv:1309.4335}},
  \href {http://dx.doi.org/10.1103/PhysRevD.88.116007}
  {\path{doi:10.1103/PhysRevD.88.116007}}.

\bibitem{Aarts:2017vrv}
G.~Aarts, E.~Seiler, D.~Sexty, I.-O. Stamatescu, {Complex Langevin dynamics and
  zeroes of the fermion determinant}, JHEP 05 (2017) 044, [Erratum:
  JHEP01,128(2018)].
\newblock \href {http://arxiv.org/abs/1701.02322} {\path{arXiv:1701.02322}},
  \href {http://dx.doi.org/10.1007/JHEP05(2017)044, 10.1007/JHEP01(2018)128}
  {\path{doi:10.1007/JHEP05(2017)044, 10.1007/JHEP01(2018)128}}.

\bibitem{Greensite:2014cxa}
J.~Greensite, {Comparison of complex Langevin and mean field methods applied to
  effective Polyakov line models}, Phys. Rev. D90~(11) (2014) 114507.
\newblock \href {http://arxiv.org/abs/1406.4558} {\path{arXiv:1406.4558}},
  \href {http://dx.doi.org/10.1103/PhysRevD.90.114507}
  {\path{doi:10.1103/PhysRevD.90.114507}}.

\bibitem{altland_simons_2010}
A.~Altland, B.~D. Simons, Condensed Matter Field Theory, 2nd Edition, Cambridge
  University Press, 2010.
\newblock \href {http://dx.doi.org/10.1017/CBO9780511789984}
  {\path{doi:10.1017/CBO9780511789984}}.

\bibitem{Gibbons:1978ac}
G.~W. Gibbons, S.~W. Hawking, M.~J. Perry, {Path Integrals and the
  Indefiniteness of the Gravitational Action}, Nucl. Phys. B138 (1978)
  141--150.
\newblock \href {http://dx.doi.org/10.1016/0550-3213(78)90161-X}
  {\path{doi:10.1016/0550-3213(78)90161-X}}.

\bibitem{Greensite:1983yc}
J.~Greensite, M.~B. Halpern, {Stabilizing bottomless action theories}, Nucl.
  Phys. B242 (1984) 167--188.
\newblock \href {http://dx.doi.org/10.1016/0550-3213(84)90138-X}
  {\path{doi:10.1016/0550-3213(84)90138-X}}.

\bibitem{Moler}
S.~I. Davis, R.~R. Ullah, C.~Adamo, C.~A. Watson, J.~R. Kirtley, M.~R. Beasley,
  S.~A. Kivelson, K.~A. Moler, Spatially modulated susceptibility in thin film
  ${\mathrm{la}}_{2\ensuremath{-}x}{\mathrm{ba}}_{x}{\mathrm{cuo}}_{4}$, Phys.
  Rev. B 98 (2018) 014506.

\end{thebibliography}

\end{document}